\documentclass[amssymb,prb,twocolumn,superscriptaddress,floats,showkeys,amsmath]{revtex4-1}

\usepackage{graphicx}
\usepackage{epstopdf}
\usepackage{hyperref} 
\hypersetup{colorlinks,citecolor=blue, filecolor=blue ,linkcolor=blue , urlcolor=blue, pdftex}
\usepackage{multirow}

\usepackage[labeled,resetlabels]{multibib}

\newcites{SI}{SupI}

\def \FUW{Institute of Experimental Physics, Faculty of Physics, University of Warsaw, ul. Pasteura 5, 02-093 Warsaw, Poland}
\def \LNCMI{Laboratoire National des Champs Magn\'etiques Intenses, CNRS-UGA-UPS-INSA-EMFL, 25, avenue des Martyrs, 38042 Grenoble, France} 
\def \Brno{Central European Institute of Technology, Brno University of Technology,  Purky\v{n}ova 656/123, 612 00 Brno, Czech Republic}
\def \Wroclaw{Department of Experimental Physics, Wrocław University of Science and Technology, ul. Wybrzeże Wyspiańskiego 27, 50-370 Wrocław, Poland}
\def \Kenji{Research Center for Functional Materials, National Institute for Materials Science, 1-1 Namiki, Tsukuba 305-0044, Japan}
\def \Takashi{International Center for Materials Nanoarchitectonics, National Institute for Materials Science, 1-1 Namiki, Tsukuba 305-0044, Japan}
\def \Tomek {Department of Semiconductor Materials Engineering, Wrocław University of Science and Technology, Wybrzeże Wyspiańskiego 27, 50-370 Wrocław, Poland}

\begin{document}

\title{Excitonic complexes in $n$-doped WS$_2$  monolayer}

\author{M. Zinkiewicz}\email{malgorzata.zinkiewicz@fuw.edu.pl}\affiliation{\FUW}
\author{T. Wo\'zniak}\affiliation{\Tomek}
\author{T. Kazimierczuk}\affiliation{\FUW}
\author{P. Kapu\'sci\'nski}\affiliation{\LNCMI}\affiliation{\Wroclaw{}}
\author{K.~Oreszczuk}\affiliation{\FUW}
\author{M.~Grzeszczyk}\affiliation{\FUW}
\author{M. Bartos}\affiliation{\LNCMI}\affiliation{\Brno}
\author{K. Nogajewski}\affiliation{\FUW}
\author{K.~Watanabe}\affiliation{\Kenji}
\author{T. Taniguchi}\affiliation{\Takashi}
\author{C. Faugeras}\affiliation{\LNCMI}
\author{P. Kossacki}\affiliation{\FUW}
\author{M. Potemski}\affiliation{\FUW}\affiliation{\LNCMI}
\author{A. Babi\'nski}\affiliation{\FUW}
\author{M. R. Molas}\email{maciej.molas@fuw.edu.pl}\affiliation{\FUW}

\begin{abstract}
We investigate the origin of emission lines apparent in the low-temperature photoluminescence spectra of $n$-doped WS$_2$ monolayer embedded in hexagonal BN layers using external magnetic fields and first-principles calculations. Apart from the neutral A exciton line, all observed emission lines are related to the negatively charged excitons. Consequently, we identify emissions due to both the bright (singlet and triplet) and dark (spin- and momentum-forbidden) negative trions as well as the phonon replicas of the latter optically-inactive complexes. The semi-dark trions and negative biexcitons are distinguished. Based on their experimentally extracted and theoretically calculated $g$-factors, we identify three distinct families of emissions due to exciton complexes in WS$_2$: bright, intravalley and intervalley dark. The $g$-factors of the spin-split subbands in both the conduction and valence bands are also determined.
\end{abstract}

\maketitle

Monolayers (MLs) of semiconducting transition metal dichalcogenides (S-TMDs) MX$_2$ where M=Mo or W and X=S, Se or Te, are direct band gap semiconductors with the minima of the conduction band (CB) and maxima of the valence band (VB) located at the inequivalent K$^\pm$ points of their hexagonal Brillouin zone (BZ).\cite{Koperski2017, Wang2018} The strong spin-orbit interaction and lack of inversion symmetry result in the splitting of the VB ($\Delta_v$) and the CB ($\Delta_c$) extrema. While the former splitting is of the order of a few hundreds of meV, the latter equals only few tens of meV and can be positive or negative.\cite{Kormanyos2015} Consequently, two subgroups of MLs can be distinguished: $bright$ (the excitonic ground state is optically active or bright) formed by MoSe$_2$ and MoTe$_2$,\cite{Molas2017, Koperski2017, Robert2020} and $darkish$ (the excitonic ground state is optically inactive or dark) composed of MoS$_2$, WS$_2$, and WSe$_2$.\cite{Molas2017, Zhang2017, Robert2020}

Dark excitons in S-TMD ML can be divided into two subgroups because of the distinct origins of their optical inactivity, $i.e.$ intravalley spin-forbidden and intervalley momentum-forbidden complexes, which can not recombine optically due to the spin and momentum conservation rule for excitons. Dark excitonic complexes can be also characterised by their net charge. Both neutral and charged dark excitons exist, which are bound electron-hole ($e$-$h$) pairs and the bound $e$-$h$ pairs with an extra carrier (an electron or a hole), respectively.

	\begin{figure}[!t]
	\centering
	\includegraphics[width=0.75\linewidth]{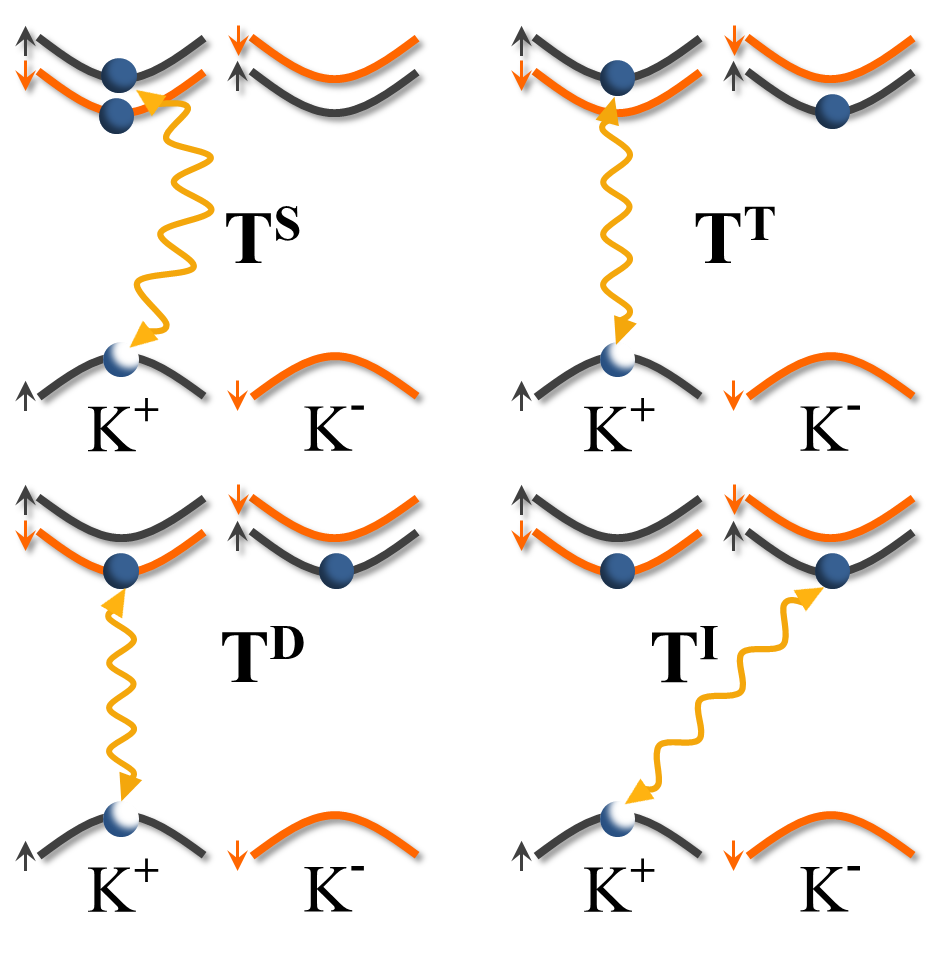}%
	\caption{Schematic illustration of possible spin configurations for negatively charged excitons formed in the vicinity of so-called A exciton. T$^\textrm{S}$ and T$^\textrm{T}$ correspond to the bright singlet and triplet trions, while T$^\textrm{D}$ and T$^\textrm{I}$ represent the dark intravalley and intervalley complexes, respectively. Note that we draw only complexes for which a hole is located at the K$^+$ point of the BZ.}
	\label{fig:scheme_trions}
\end{figure}

In this work we investigate the low-temperature optical response of high-quality $n$-doped WS$_2$ ML encapsulated hexagonal BN (hBN) flakes using photoluminescence (PL) spectroscopy in external magnetic fields. All emission lines, observed in the PL spectrum, are due to both the bright (singlet and triplet) and dark (spin- and momentum-forbidden) negative trions as well as the phonon replicas of the latter optically-inactive complexes. Moreover, the semi-dark trions and negative biexcitons are distinguished. Magneto-PL measurements accompanied with first-principles calculations allow us to extract the $g$-factors of all transitions as well as of the spin-split subbands in both the conduction and the valence bands.

The negatively charged exciton or negative trion is a three-particle complex composed of an $e$-$h$ pair and an excess electron. There are four negative trions in W-based darkish MLs (WS$_2$ or WSe$_2$), $i.e.$ two bright and two dark states, see Fig.~\ref{fig:scheme_trions}. These states can be formed in both the K$^+$ and K$^-$ valleys (taking into account the location of a hole), which leads to two possible configurations of a given complex. Due to the spin conservation rule for S-TMD MLs, the bright (optically active) negative trion may be found in both the intravalley singlet (T$^\textrm{S}$) and intervalley triplet (T$^\textrm{T}$) states. They involve correspondingly two electrons from the same valley whereas the triplet trion comprises two electrons from different valleys. For the dark (optically inactive) negative trions, the corresponding electrons are located in different valleys and are characterized by the antiparallel alignment of their spins. This configuration leads to two complexes, depending on the electron involved in recombination process: intravalley spin- (T$^\textrm{D}$) and intervalley momentum-forbidden (T$^\textrm{I}$), which cannot recombine optically because of the spin and momentum conservation, respectively.

\begin{figure}[t]
		\centering
		\includegraphics[width=1\linewidth]{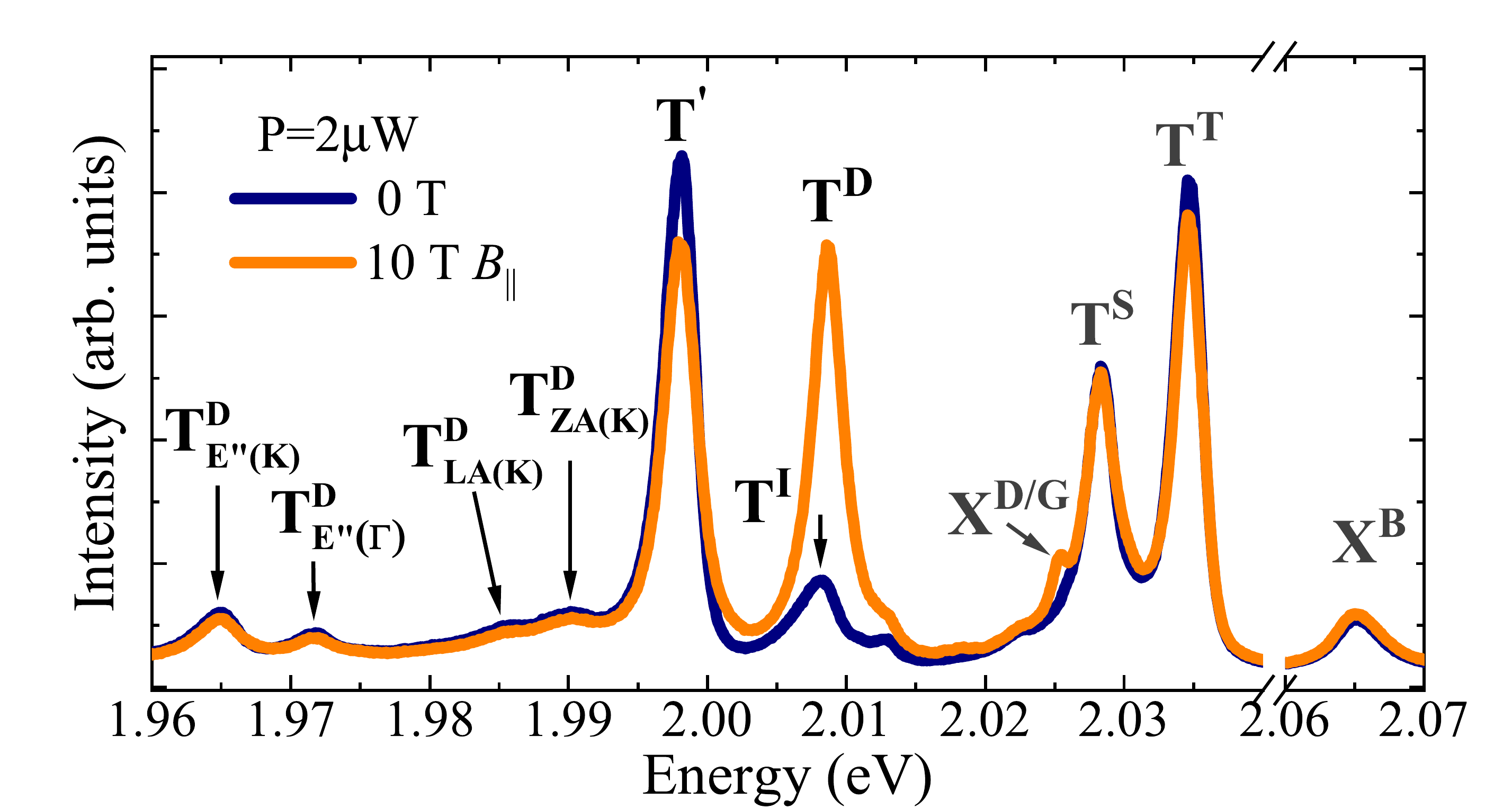}%
		\caption{Low-temperature PL spectra measured on an $n$-doped WS$_2$ ML encapsulated in hBN flakes at zero magnetic field and in the in-plane magnetic field of $B_{||}$=10~T.}
		\label{fig:lowPL}
\end{figure}

In order to investigate the negatively charged complexes in our WS$_2$ ML, we measured its low temperature ($T$=5~K) PL spectrum at zero magnetic field and in the in-plane magnetic field of 10~T, see Fig.~\ref{fig:lowPL}. It is well established that the application of an in-plane magnetic field ($B_{||}$) results in the mixing of the bright and dark excitons, which becomes apparent in the optical activation of spin-forbidden dark complexes~\cite{Slobodeniuk2016, Molas2017, Zhang2017, Molas2019Dark, Lu2019,Robert2020,Zinkiewicz2020}. The zero-field PL spectrum is composed of several emission lines. Based on the previous reports~\cite{Molas2017, Vaclavkova2018, nagler2018, Jadczak2019, Paur2019}, three peaks can be assigned unquestionably to the bright neutral exciton (X$^\textrm{B}$) and to two bright negatively charged excitons, $i.e.$ singlet (T$^\textrm{S}$) and triplet (T$^\textrm{T}$), formed in the vicinity of the optical band gap (A exciton). Two additional lines, labelled X$^\textrm{D/G}$ and T$^\textrm{D}$ become apparent in $B_{||}$=10~T (see Fig.~\ref{fig:lowPL}). As it was recently reported~\cite{Zinkiewicz2020}, the X$^\textrm{D/G}$ peak corresponds to the dark and grey states of the neutral exciton, while the T$^\textrm{D}$ peak is related to the dark state of the negative trion. The PL spectra, shown in Fig.~\ref{fig:lowPL}, comprise also several lines at lower energies, which are denoted as T$^\textrm{I}$, T', T$^\textrm{D}_\textrm{ZA(K)}$, T$^\textrm{D}_\textrm{LA(K)}$, T$^\textrm{D}_\textrm{E"(K)}$, and T$^\textrm{D}_{\textrm{E"}(\Gamma)}$. Increasing the excitation power leads to the appearance of an additional emission, labelled XX$^-$. Those lines have not been reported so far in WS$_2$ MLs and the following is dedicated to their identification.

\begin{figure}[t]
		\centering
		\includegraphics[width=1\linewidth]{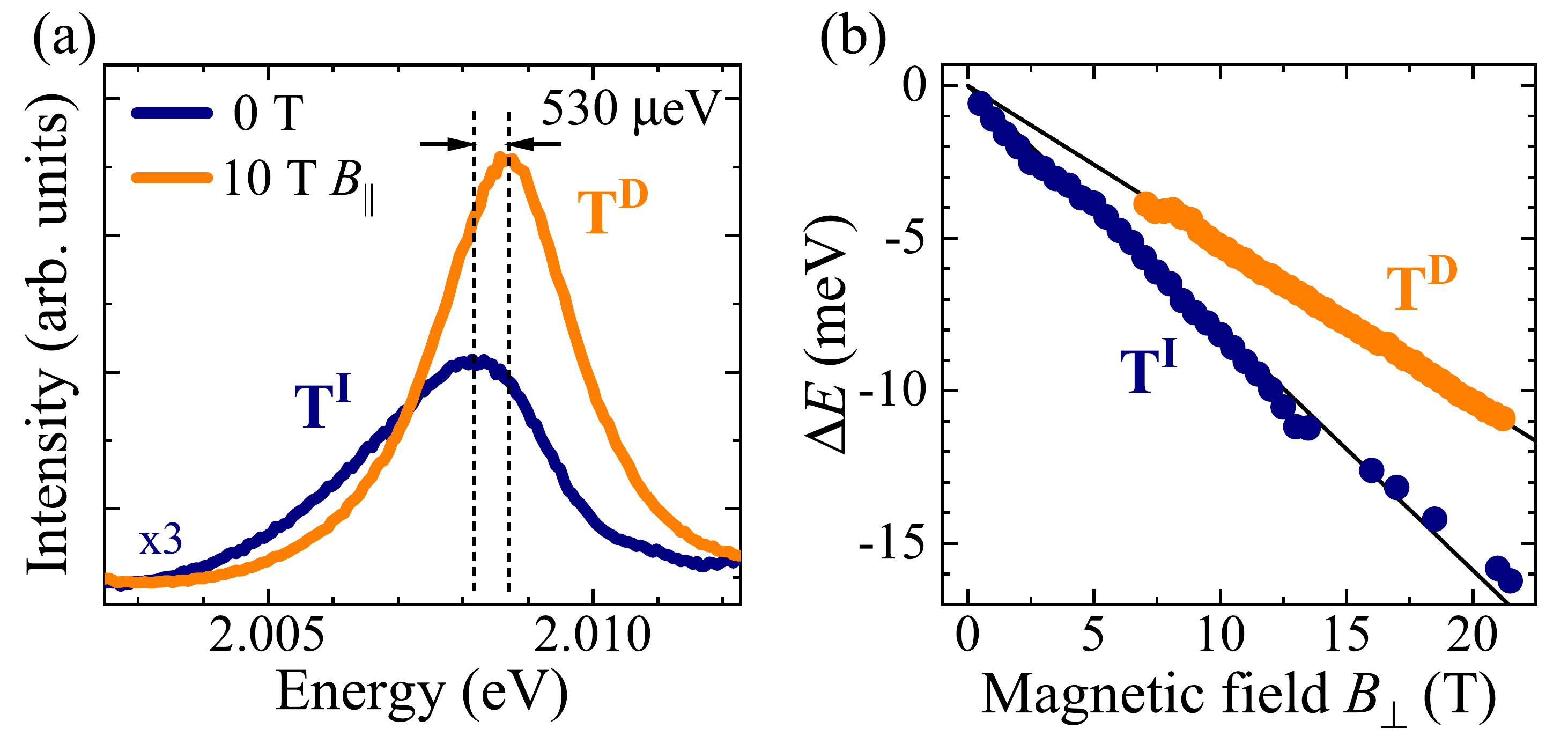}%
		\caption{(a) Low-temperature emission due to the intravalley T$^\textrm{D}$ and intervalley T$^\textrm{I}$ dark trions measured on an $n$-doped WS$_2$ ML encapsulated in hBN flakes at zero magnetic field and in the in-plane magnetic field of $B_{||}$=10~T. (b) Energy separation between the two circularly-polarized split components of the T$^\textrm{D}$ and T$^\textrm{I}$ transitions as a function of the out-of-plane magnetic field $B_\perp$. The solid lines represent fits according to the equation described in the text. Note the measurements were performed in the tilted configuration of the magnetic field direction in the respect to the ML plane (see Supplementary Information for details).}
		\label{fig:TDvsTI}
\end{figure}

	\begin{figure*}[!t]
		\centering
		\includegraphics[width=1\linewidth]{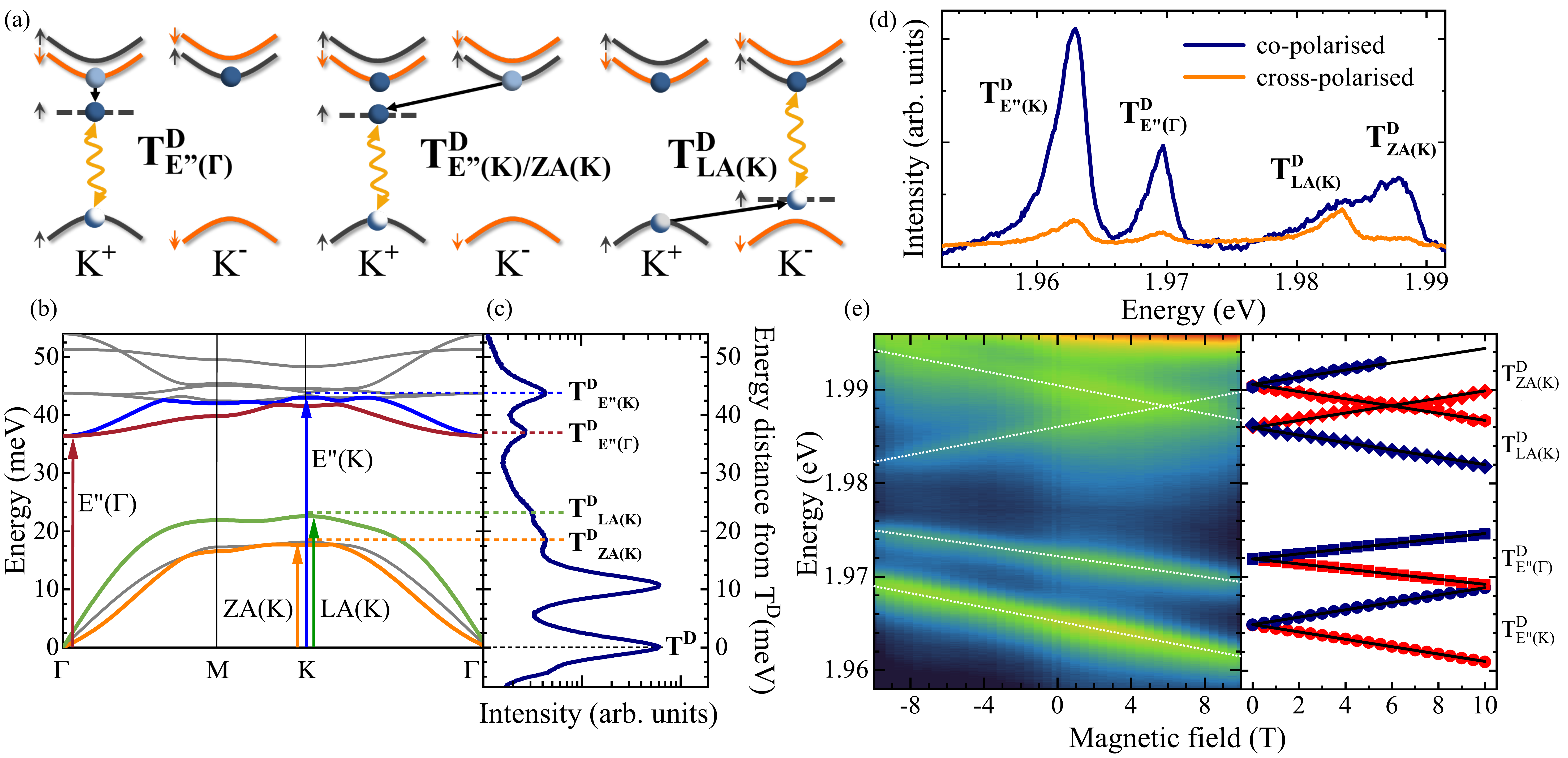}%
		\caption{(a) Schematic illustration of possible recombination pathways of dark trions assisted by the emission of optical (E") and acoustic (ZA, LA) phonons from the K or $\Gamma$ points of the BZ, which give rise to the PL of so-called phonon replicas. The black solid lines represent the phonon emission, which transfer an electron or a hole from the real subband in the CB or VB to the virtual state denoted by a dashed horizontal line. Only complexes for which a hole is located at the K$^+$ point of the BZ are drawn. (b) The calculated phonon dispersion of WS$_2$ ML. The dispersions of the pertinent phonon modes are indicated by colour curves, while for the others it is represented by gray curves. (c) Low-temperature PL spectrum due to the intravalley T$^\textrm{D}$ dark trion and its phonon replicas measured on the studied WS$_2$ ML at $B_{||}$=10~T. Note that the energy axis in panel (c) is relative, $i.e.$ in reference to the T$^\textrm{D}$ emission. (d) Helicity resolved low-temperature PL spectrum with the emission lines related to phonon replicas of the studied ML at zero magnetic field under circularly polarized excitation with the T$^\textrm{D}$ energy. (e) (left panel) False-color PL map as a function of out-of-plane magnetic field ($B_\perp$). Note that the positive and negative values of magnetic fields correspond to $\sigma^\pm$ polarizations of detection. White dashed lines superimposed on the observed transitions are guides to the eyes. (right panel) Transition energies of the $\sigma^\pm$ (red/blue points) components of the studied line related to phonon replicas as a function of the out-of-plane magnetic field. The solid lines represent fits according to the equation described in the text.}
		\label{fig:replicas}
	\end{figure*}

As can be appreciated in Fig.~\ref{fig:scheme_trions}, both the intravalley spin-forbidden and intervalley momentum-forbidden negative trions share the same carrier configuration. The difference arises from their recombination pathway. While the T$^\textrm{D}$ complex involves recombination of an $e$-$h$ pair from the same K$^\pm$ point, for the T$^\textrm{I}$ trion, an electron and a hole from neighbouring K$^\pm$ valleys recombine. The T$^\textrm{D}$ can be identified with magnetic brightening experiments with a $B_{||}$ field, see Fig.~\ref{fig:TDvsTI}(a). This effect was investigated in details in Ref.~\citenum{Zinkiewicz2020}. The observation and assignment of the emission due to the intervalley momentum-forbidden negative trion is more striking. Recently, similar emission related to the momentum-forbidden dark neutral exciton was reported in the WSe$_2$ ML~\cite{he2020valley, Liu2020}. It was demonstrated that at zero magnetic field the intensity of the momentum-forbidden emission is much smaller as compared to the corresponding spin-forbidden one. In our case, the T$^\textrm{I}$ line dominates at $B$=0~T, while the T$^\textrm{D}$ one can be only observed because of the brightening effect. As the energy difference between those complexes is only of about 530~$\mu$eV, the optical emission of the T$^\textrm{I}$ line through the Auger processes~\cite{Danovich_2016} or by emission of optical phonons can be excluded~\cite{he2020valley}. Similar to the case of the indirect band gap in thin layers of WS$_2$~\cite{Molas2017nano}, the optical recombination of intervalley dark trion can be allowed due to defect states, which may provide momentum conservation during recombination. The origin of the T$^\textrm{D}$-T$^\textrm{I}$ energy splitting is not clear as both the intervalley and intravalley dark trions share the same carrier configuration (see Fig.~\ref{fig:scheme_trions}). We believe that this splitting arises from higher-order processes, description of which is beyond the scope of our work. The similar energy separation between the intervalley and intravalley dark neutral excitons in WSe$_2$ ML was reported to be on the order of 10~meV and is ascribed to a short-range electron-hole exchange interaction~\cite{he2020valley,Liu2020}. 

Upon application of an out-of-plane magnetic field ($B_\perp$), excitonic transitions split into two circularly polarized components ($\sigma^\pm$). Their energy separation $\Delta E(B)=E_{\sigma^+}-E_{\sigma^-}$ can be expressed as $\Delta E(B)=g \mu_B B_\perp$, where $g$ denotes the $g$-factor of the considered excitonic complex and $\mu_B$ is the Bohr magneton.
The magnetic field evolution of the $\Delta E$ with linear fits to experimental data for both the T$^\textrm{D}$ and T$^\textrm{I}$ excitons is shown in Fig.~\ref{fig:TDvsTI}(b). Linear fits to the experimental data are also presented in the Figure. The resulting $g$-factors of the T$^\textrm{D}$ and T$^\textrm{I}$ are equal to \mbox{--8.9} and \mbox{--13.7}, respectively. The former value is consistent with our recent result reported in Ref.~\citenum{Zinkiewicz2020}, while the latter one is very similar to the $g$-factor reported for intervalley dark complexes in WSe$_2$ MLs~\cite{he2020valley,Liu2020}. 

The identification of four lines apparent in the lowest energy range of the PL spectrum (see Fig.~\ref{fig:lowPL}) will be addressed in the following. One of the possibilities to fulfil the spin and momentum conservation during optical recombination of the dark trions is phonon emission. Fig. \ref{fig:replicas}(a) shows a schematic illustration of possible recombination pathways of dark negative trions involving phonon emission. The phonon-assisted processes give rise to so-called phonon replicas of dark excitons in WSe$_2$ MLs~\cite{Li2019Momentum, Li2019Replica, Liu2019Replica, he2020valley, Liu2020, Robert2020gfactor}. Due to the symmetry and the momentum of a given phonon, it may lead to the transfer of a carrier (an electron or a hole) to a virtual state with a spin-flip or a valley-flip. 

The phonon replicas of the dark trion should be redshifted from it by the phonon energies, with the redshift corresponding to the phonon emission. The calculated phonon dispersion can be therefore compared to the low-temperature PL spectrum of the WS$_2$ ML as presented in Figs~\ref{fig:replicas}(b) and (c). We found that the extracted relative energies of phonon replicas from the T$^\textrm{D}$ line are in good agreement with the corresponding theoretical phonon energies. To confirm the assignment of phonons shown in Fig.~\ref{fig:replicas}(a), we analysed their symmetries following the group theory considerations and irreducible representations (IR) notation from Ref.~\cite{He2020}, where WSe$_2$ ML, of the same symmetry as WS$_2$ ML, has been studied. The intravalley spin-flip process of an electron can only be assisted by a phonon, which transforms like the IR $\Gamma_5$, $i.e.$ E"($\Gamma$). Additionally, comparing the measured redshift (37 meV) and the calculated phonon energy (36.4 meV), the phonon replica observed at about 1.97~eV can be identified as T$^\textrm{D}_\textrm{E''($\Gamma$)}$. The other replicas, which involve the momentum-flip processes (the spin of the electrons is conserved at the same time), must be induced by phonons from K point. It is possible to transfer an electron (hole)  between K$^{\pm}$ valleys with emission of E"(K) (LA(K)) phonons, as they transform according to IR K$_3$ (K$_1$). Their calculated energies (43.0 meV and 22.6 meV) agree well with the measured redshifts (44 meV and 23 meV). Therefore, we label the replicas apparent at about 1.965~eV and 1.986~eV as T$^\textrm{D}_\textrm{LA(K)}$ and T$^\textrm{D}_\textrm{E''(K)}$. The red shift of the replica at around 1.99~eV (18 meV) is close to calculated energies of TA(K) and ZA(K) phonons (18.2 and 17.7 meV). The former one should induce a spin flip, while the latter one preserves this symmetry and couples to a spin conserving transition. Due to the extracted $g$-factor value for this replica (discussed in the next paragraph), we label it as T$^\textrm{D}_\textrm{ZA(K)}$.

To confirm the assignment of the phonon replicas, we analyse their polarization properties under circular polarized excitation and their evolution when applying an out-of-plane magnetic field. Fig.~\ref{fig:replicas}(d) shows the helicity-resolved PL under circularly polarized excitation with the T$^\textrm{S}$ energy. After formation of the T$^\textrm{S}$ complex at the K$^\pm$ point using the $\sigma^\pm$ polarization, an electron from the top subband of the CB at K$^\pm$ point relaxes to the bottom one at the opposite K$^\mp$ point, which leads to the formation of the dark trion (see Fig.~\ref{fig:scheme_trions}). As a result of spin- and momentum-flip processes of electrons in the CB, three replicas (T$^\textrm{D}_\textrm{E''($\Gamma$)}$, T$^\textrm{D}_\textrm{E''(K)}$, T$^\textrm{D}_\textrm{ZA(K)}$) are characterized by large conservation of excitation helicity in emission. For the T$^\textrm{D}_\textrm{LA(K)}$ line, the opposite behaviour should be present, as the emission occurs at the valley which is opposite to excitation one due to the momentum-flip of a hole in the VB. However, that replica demonstrates almost zero preservation of excitation helicity, which may be related to the scattering processes of carriers~\cite{Singh2016}. These polarization properties also affect the PL spectra measured in $B_\perp$ field. As can be seen in Fig.~\ref{fig:replicas}(e), the studied lines are characterized not only by different magnitudes of their field-induced shifts but also by the sign. To extract their $g$-factors, we fitted our experimental results using formula $E_{\sigma^\pm}(B)=E_0 \pm 1/2 g \mu_B B_\perp$, where $E_0$ is the emission energy at zero field. The obtained $g$-factors of -13.6, -13.3, and +13.6 for T$^\textrm{D}_\textrm{E''(K)}$, T$^\textrm{D}_\textrm{ZA(K)}$, and T$^\textrm{D}_\textrm{LA(K)}$, respectively, are consistent with the T$^\textrm{I}$ one (-13.7). Note that the sign of the T$^\textrm{D}_\textrm{LA(K)}$ $g$-factor is opposite, which is an indication of the intervalley transfer of hole, whereas the $g$-factor of T$^\textrm{D}_\textrm{E''($\Gamma$)}$ of -8.9 is identical to the value obtained for T$^\textrm{D}$. 

	\begin{figure}[!t]
	\centering
	\includegraphics[width=1\linewidth]{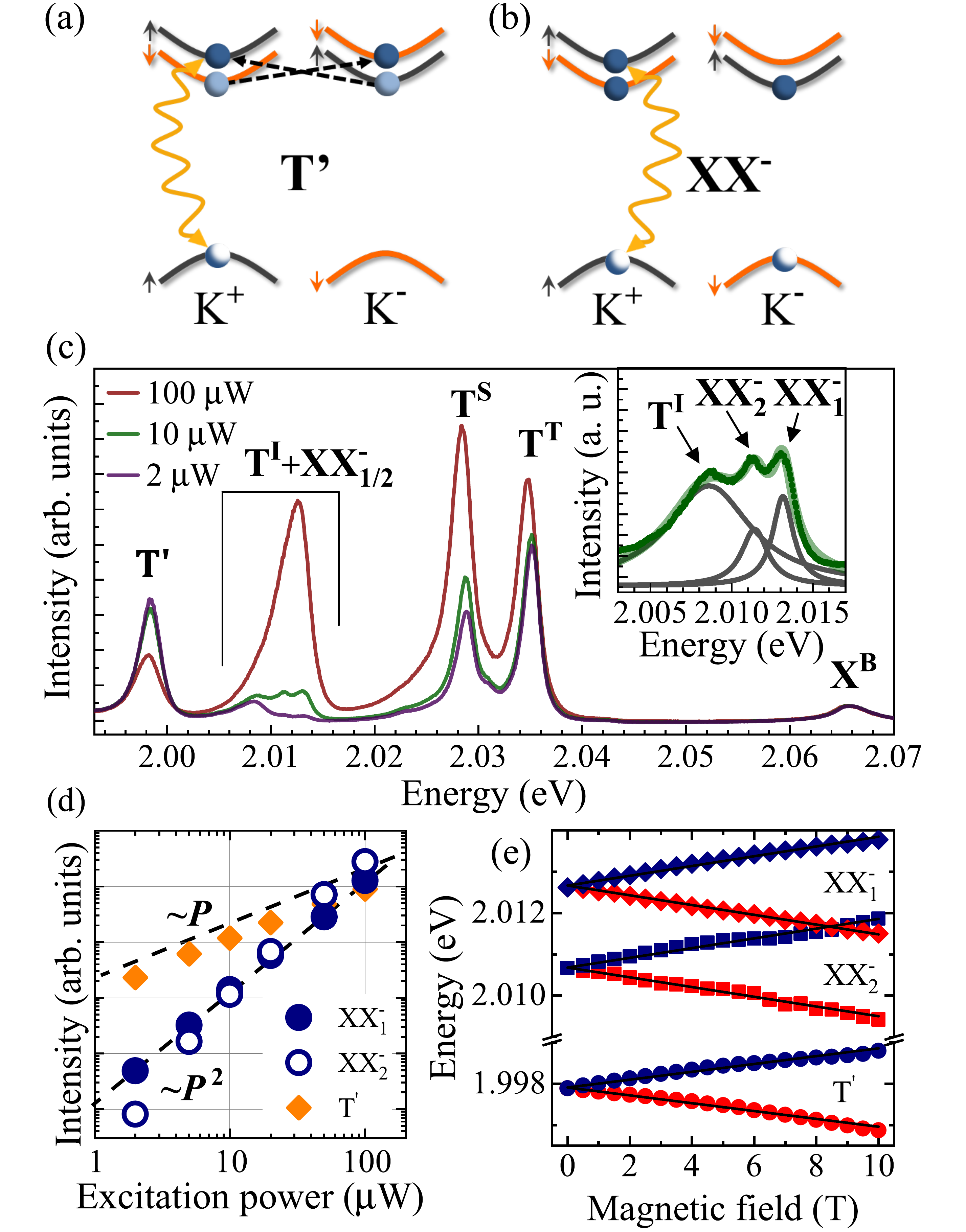}%
	\caption{Schematic illustration of a possible recombination pathway of (a) semi-dark trions made optically active due to the electron-electron ($e$-$e$) scattering and (b) negative biexcitons. The black dashed lines represent $e$-$e$ scattering, which transfers an electron from the lower subband of the CB at the K$^\pm$ point to the corresponding higher subband at the opposite K$^\mp$ point (c) Power dependence of the low-temperature PL spectra of the WS$_2$ ML. The intensities of the PL spectra are normalized by the X$^\textrm{B}$ intensity. The inset displays the PL spectrum under 10 $\mu$W excitation due to the T$^\textrm{I}$, XX$^-_1$, and XX$^-_2$ lines deconvoluted using Lorentzian function. (d) The intensity evolution of the emission features with excitation power. The dashed black line indicates the linear and quadratic behaviours as a guide to the eye. (e) Transition energies of the $\sigma^\pm$ (red/blue points) components of the T', XX$^-_1$, and XX$^-_2$ lines as a function of the out-of-plane magnetic field. The solid lines represent fits according to the equation described in the text. }
	\label{fig:semi-dark}
\end{figure}

One of the most pronounced features observed in the PL spectrum is the T' line, which intensity is comparable to the T$^\textrm{S}$ and T$^\textrm{T}$ ones, see Fig.~\ref{fig:lowPL}. Although the emission line was previously observed many times in the low-temperature PL spectra of both the WSe$_2$ and WS$_2$ MLs~\cite{Paur2019, Li2019Trion, Li2019Replica, Li2019Momentum, Liu2019, Liu2019Replica, he2020valley, Liu2020}, its origin is not well defined. We ascribe the T' line to the recombination of the dark trion made optically active due to the electron-electron ($e$-$e$) scattering~\cite{Danovich2017}, and name it a semi-dark trion. We note that the corresponding semi-dark trion has recently been reported in WSe$_2$ ML ~\cite{Tu2019}. The initial state of the T' line is dark, the same as for both the T$^\textrm{D}$ and T$^\textrm{I}$ trions, compare Figs.~\ref{fig:semi-dark}(a) and \ref{fig:scheme_trions}. Because of the intervalley $e$-$e$ scattering, an electron located at the lower CB subband in the K$^\pm$ point is transferred to the higher lying CB subband in the opposite K$\mp$ valley, which results in the optical recombination of an $e$-$h$ pair.

The last studied excitonic complex is a negative biexciton, denoted as XX$^-$. Its formation is possible due to the long lifetime of dark trions (see Fig.~\ref{fig:semi-dark}(b)), which was reported to be close to 0.5~ns~\cite{Zinkiewicz2020}. The emission related to negative biexcitons is only apparent at high excitation power and it is characterized by two lines, labelled XX$^-_1$ and XX$^-_2$, see Fig.~\ref{fig:semi-dark}(c). The XX$^-_1$--XX$^-_2$ energy separation being about 2~meV is very similar to the energy separation reported for two neutral biexcitons (2.5~meV) in WSe$_2$ ML~\cite{Barbone2018}. Surprisingly, as two lines of the neutral biexcitons can be explained in terms of two possible carrier configurations, the appearance of the XX$^-_1$ and XX$^-_2$ lines is not clear (there is a single possible configuration, see Fig.~\ref{fig:semi-dark}(b)) which requires more sophisticated theoretical analysis.

\begin{table}[t]
\caption{\label{tab:exciton_g}
Experimental ($g^{exp}$) and theoretical ($g^{calc}$) $g$-factors of investigated emission lines. $\Delta L_{K^+}$ and $\Delta \Sigma_{K^+}$ represent orbital and spin contributions to $g_{calc}$ at K$^+$, respectively. The helicity of the given transition occurring at K$^+$ valley, denoted as $P$, is shown in the last column.  }
\begin{tabular}{cccccc}
    & $g^{exp}$	& $g^{calc}$ & $\Delta L$ & $\Delta \Sigma$ & $P$	\\
\hline		
X$^\textrm{B}$  & -3.5  & \multirow{6}{*}{-3.56} & \multirow{6}{*}{-1.78} & \multirow{6}{*}{0}  & \multirow{6}{*}{$\sigma+$} \\
T$^\textrm{S}$  & -4.0  \\
T$^\textrm{T}$                          & -3.9 \\
T'                                      & -3.3 \\
XX${}^-_1$                              & -4.1  \\
XX${}^-_2$                              & -4.1  \\
\hline
T$^\textrm{D}$ 	                        & -8.9  & \multirow{2}{*}{-8.73}  & \multirow{2}{*}{-2.37} & \multirow{2}{*}{-2} & \multirow{2}{*}{$\sigma+$} \\
T$^\textrm{D}_\textrm{E''($\Gamma$)}$   & -8.9  \\
\hline
T$^\textrm{I}$                          & -13.7 & \multirow{3}{*}{-12.20} & \multirow{3}{*}{-6.10} & \multirow{3}{*}{0}  & \multirow{3}{*}{$\sigma+$} \\
T$^\textrm{D}_\textrm{E''(K)}$          & -13.6 & \\
T$^\textrm{D}_\textrm{ZA(K)}$           & -13.3 & \\
\hline
T$^\textrm{D}_\textrm{LA(K)}$	        & +13.6 & +12.20 & -6.10 & 0  & $\sigma-$ \\
\hline
\end{tabular}

\end{table}

To confirm our assignment of the T', XX$^-_1$, and XX$^-_2$ lines, Fig.~\ref{fig:semi-dark}(d) presents their intensity evolution as a function of excitation power. We found that T' emission follows a linear behaviour, while the XX$^-_1$ and XX$^-_2$ peaks are characterized by superlinear evolution. These types of power dependence are typical for excitonic complexes composed of a single $e$-$h$ pair or by two $e$-$h$ pairs~\cite{Barbone2018,Chen2018,Li2018}. Note that the excitation-power evolution of other excitons is presented in Supplementary Information (SI). Additionally, we analyse their evolution in the $B_\perp$ fields, see Fig.~\ref{fig:semi-dark}(e). Using the same approach as for phonon replicas, we extracted $g$-factors of \mbox{--3.3}, \mbox{--4.1}, and \mbox{--4.1} for T', XX$^-_1$, and XX$^-_2$ lines, respectively. These values are consistent with the $g$-factors found for the bright complexes, such as X$^\textrm{B}$, T$^\textrm{S}$, T$^\textrm{T}$ (see SI for details).

The $g$-factors for the all studied excitonic complexes are summarized in Tab.~\ref{tab:exciton_g}. According to the extracted values of $g$-factors, the excitonic complexes can be arranged into three groups: (i) $g$-factors in the range \mbox{--3.3} -- \mbox{--4.1} are characteristic for bright transitions (X$^\textrm{B}$, T$^\textrm{S}$, T$^\textrm{T}$, T', XX${}^-_1$, and XX${}^-_2$); (ii) the spin-forbidden dark transitions are described by \mbox{--8.9} values of $g$-factors (T$^\textrm{D}$ and T$^\textrm{D}_\textrm{E''($\Gamma$)}$); and (iii) values of $g$-factors of about \mbox{--13.3} -- \mbox{--13.7} and \mbox{+13.6} are obtained for momentum-forbidden dark transitions (T$^\textrm{I}$, T$^\textrm{D}_\textrm{E''(K)}$, T$^\textrm{D}_\textrm{ZA(K)}$, and T$^\textrm{D}_\textrm{LA(K)}$). In order to establish the $g$-factors of a single subbands in both the CB and VB, we adapted the method proposed in Ref.~\citenum{Robert2020gfactor}. It relies on the comparison of $g$-factors related to the different excitonic complexes (see SI for details). The obtained values of the $g$-factors for higher-energy $c+1$ ($v$) and lower-energy $c$ ($v-1$) subbands in CB (VB) at the K$^+$ point are shown in Tab.~\ref{tab:band_g}. Note that the $g$-factor of $v-1$ band was calculated using the reported $g$-factor of B exciton (-4) in Ref. \citenum{Stier2016}. The extracted values demonstrate that the simple model commonly employed for calculation of the excitonic $g$-factors using additive contribution of the spin, valley, and orbital angular momenta~\cite{Aivazian2015} cannot explain the single band $g$-factors.

\begin{table}[t]
\caption{\label{tab:band_g}
Lower: Experimental ($g^{exp}_{n}$) and theoretical ($g^{calc}_{n}$) $g$-factors, orbital (L$_n$) and spin ($\Sigma_n$) angular momenta of valence (n: v-1, v) and conduction (n: c, c+1) bands at  K$^+$ point.}
\begin{tabular}{ccccc}
n     & $g^{exp}_{n}$ & $g^{calc}_{n}$ & L$_{n}$ & ${\Sigma}_{n}$    \\
\hline
v-1   & 2.81  & 2.79  & 3.79  & -1 \\
v     & 6.10  & 5.23  & 4.23  & +1 \\
c     & 0.86  & 0.87  & 1.87  & -1 \\
c+1   & 3.84  & 3.45  & 2.45  & +1 \\
\hline
\end{tabular}

\end{table}

To verify our experimental results, we calculate theoretically $g$-factors using a first principles based approach proposed in Ref.~\citenum{Wozniak2020}. In this case, first the $g$-factors of single subbands ($g^{calc}_{n}$) are calculated, which are then used to determine the $g$-factor of a given transition ($g^{calc}$). Consequently, the theoretical $g$-factor of the band $n$ ($n=v-1,v,c,c+1$) at point K$^+$ is evaluated as $g_{n,K+}^{calc} = L_{n,K+} + \Sigma_{n,K+}$, where $L_{n,K+}$ and $\Sigma_{n,K+}$ are the orbital and spin angular momenta of the bands hosting the bound $e$-$h$ pair involved in the optical recombination process, while the excess carriers do not contribute. A scissor correction to the experimental free-particle gap was applied during the evaluation of $L_{n,K+}$ \cite{Molas2019Spectrum}. The $g$-factors of studied excitons can then be expressed as $g^{calc} = \pm 2 (g_{c(+1),K+}^{calc} - g_{v(-1),K+}^{calc}) = \pm 2 (\Delta L_{K+} + \Delta \Sigma_{K+})$, where $\Delta$ denotes the difference of respective angular momenta between the $c(+1)$ and $v(-1)$ bands. The sign is determined by the polarization of the transition at K$^+$ valley, which reflects the optical selection rules. We obtain three values of $g^{calc}$: -3.56, -8.73 and $\pm$12.20, which correspond to bright, spin-forbidden, and momentum-forbidden transitions groups, respectively. The magnitudes of $g$-factors can be explained in terms of the orbital and spin contributions to the angular momenta of bands. The bright complexes involve the spin-conserving transitions, therefore their $g^{calc}$ is determined only by the orbital contribution. For the spin-forbidden complexes, the change of spin part is equal to -2 and the orbital part is also enhanced, as band $c$ is involved in the transition. The momentum-forbidden complexes involve carriers from different K valleys, which gives rise to large orbital angular momentum difference and leads to a high value of $g^{calc}$. For T$^\textrm{D}_\textrm{LA(K)}$ the $g$-factor is positive because K$^-$ valley couples to $\sigma^-$ light. As can be appreciated in Tab.~\ref{tab:band_g}, the experimental and theoretical values of $g$-factors are in very good agreement. The spread of experimentally obtained $g$-factors from -3.3 to -4.1 for bright group cannot be explained with the employed approach and requires further theoretical investigations.

We identified all emission lines apparent in the low-temperature PL spectra of $n$-doped WS$_2$ ML embedded in hexagonal BN layers using external magnetic fields. We found that the extracted $g$-factors of all transitions may be arranged in three groups revealing a nature of electron-hole recombination: bright, intravalley, and intervalley dark. We explained their signs and magnitudes with the aid of first-principles calculations. This division can open an opportunity to identify the origin of the reported so-called localized excitons in the emission spectra of the WSe$_2$ and WS$_2$ MLs exfoliated on Si/SiO$_2$ substrates. The obtained $g$-factors of the spin-split subbands in both the CB and VB are important for better understanding of the interlayer transitions in van der Waals heterostructures.

\section*{acknowledgement}
We thank A. O. Slobodeniuk, M. Bieniek, and P. E. Faria Junior for fruitful discussions. The work has been supported by the the National Science Centre, Poland (grants no. 2017/27/B/ST3/00205, 2017/27/N/ST3/01612 and 2018/31/B/ST3/02111), EU Graphene Flagship project (no. 785219), the ATOMOPTO project (TEAM programme of the Foundation for Polish Science, co-financed by the EU within the ERD-Fund), the Nano fab facility of the Institut N\'eel, CNRS UGA, and the LNCMI-CNRS, a member of the European Magnetic Field Laboratory (EMFL). The Polish participation in EMFL is supported by the DIR/WK/2018/07 grant from Polish Ministry of Science and Higher Education. T.W.~acknowledges financial support by the Polish Ministry of Science and Higher Education via the "Diamond Grant" no. D\textbackslash2015 002645. P. K. kindly acknowledges the National Science Centre, Poland (grant no. 2016/23/G/ST3/04114) for financial support for his PhD. M. B. acknowledges the financial support from the ERC under the European Union's Horizon 2020 research and innovation programme (GA no. 714850) and of the Ministry of Education, Youth and Sports of the Czech Republic under the project CEITEC 2020 (Grant No. LQ1601). K.W. and T.T. acknowledge support from the Elemental Strategy Initiative conducted by the MEXT, Japan, (grant no. JPMXP0112101001), JSPS KAKENHI (grant no. JP20H00354), and the CREST (JPMJCR15F3), JST. The calculations were carried out with the support of the Interdisciplinary Centre for Mathematical and Computational Modelling (ICM), University of Warsaw, under grant no GB69-17. 

\bibliographystyle{apsrev4-1}
\bibliography{Dark_WS2}

\begin{thebibliography}{35}%
\makeatletter
\providecommand \@ifxundefined [1]{%
 \@ifx{#1\undefined}
}%
\providecommand \@ifnum [1]{%
 \ifnum #1\expandafter \@firstoftwo
 \else \expandafter \@secondoftwo
 \fi
}%
\providecommand \@ifx [1]{%
 \ifx #1\expandafter \@firstoftwo
 \else \expandafter \@secondoftwo
 \fi
}%
\providecommand \natexlab [1]{#1}%
\providecommand \enquote  [1]{``#1''}%
\providecommand \bibnamefont  [1]{#1}%
\providecommand \bibfnamefont [1]{#1}%
\providecommand \citenamefont [1]{#1}%
\providecommand \href@noop [0]{\@secondoftwo}%
\providecommand \href [0]{\begingroup \@sanitize@url \@href}%
\providecommand \@href[1]{\@@startlink{#1}\@@href}%
\providecommand \@@href[1]{\endgroup#1\@@endlink}%
\providecommand \@sanitize@url [0]{\catcode `\\12\catcode `\$12\catcode
  `\&12\catcode `\#12\catcode `\^12\catcode `\_12\catcode `\%12\relax}%
\providecommand \@@startlink[1]{}%
\providecommand \@@endlink[0]{}%
\providecommand \url  [0]{\begingroup\@sanitize@url \@url }%
\providecommand \@url [1]{\endgroup\@href {#1}{\urlprefix }}%
\providecommand \urlprefix  [0]{URL }%
\providecommand \Eprint [0]{\href }%
\providecommand \doibase [0]{http://dx.doi.org/}%
\providecommand \selectlanguage [0]{\@gobble}%
\providecommand \bibinfo  [0]{\@secondoftwo}%
\providecommand \bibfield  [0]{\@secondoftwo}%
\providecommand \translation [1]{[#1]}%
\providecommand \BibitemOpen [0]{}%
\providecommand \bibitemStop [0]{}%
\providecommand \bibitemNoStop [0]{.\EOS\space}%
\providecommand \EOS [0]{\spacefactor3000\relax}%
\providecommand \BibitemShut  [1]{\csname bibitem#1\endcsname}%
\let\auto@bib@innerbib\@empty
\bibitem [{\citenamefont {Koperski}\ \emph {et~al.}(2017)\citenamefont
  {Koperski}, \citenamefont {Molas}, \citenamefont {Arora}, \citenamefont
  {Nogajewski}, \citenamefont {Slobodeniuk}, \citenamefont {Faugeras},\ and\
  \citenamefont {M.}}]{Koperski2017}%
  \BibitemOpen
  \bibfield  {author} {\bibinfo {author} {\bibfnamefont {M.}~\bibnamefont
  {Koperski}}, \bibinfo {author} {\bibfnamefont {M.~R.}\ \bibnamefont {Molas}},
  \bibinfo {author} {\bibfnamefont {A.}~\bibnamefont {Arora}}, \bibinfo
  {author} {\bibfnamefont {K.}~\bibnamefont {Nogajewski}}, \bibinfo {author}
  {\bibfnamefont {A.~O.}\ \bibnamefont {Slobodeniuk}}, \bibinfo {author}
  {\bibfnamefont {C.}~\bibnamefont {Faugeras}}, \ and\ \bibinfo {author}
  {\bibfnamefont {P.}~\bibnamefont {M.}},\ }\href {\doibase
  10.1515/nanoph-2016-0165} {\bibfield  {journal} {\bibinfo  {journal}
  {Nanophotonics}\ }\textbf {\bibinfo {volume} {6}},\ \bibinfo {pages} {1289}
  (\bibinfo {year} {2017})}\BibitemShut {NoStop}%
\bibitem [{\citenamefont {Wang}\ \emph {et~al.}(2018)\citenamefont {Wang},
  \citenamefont {Chernikov}, \citenamefont {Glazov}, \citenamefont {Heinz},
  \citenamefont {Marie}, \citenamefont {Amand},\ and\ \citenamefont
  {Urbaszek}}]{Wang2018}%
  \BibitemOpen
  \bibfield  {author} {\bibinfo {author} {\bibfnamefont {G.}~\bibnamefont
  {Wang}}, \bibinfo {author} {\bibfnamefont {A.}~\bibnamefont {Chernikov}},
  \bibinfo {author} {\bibfnamefont {M.~M.}\ \bibnamefont {Glazov}}, \bibinfo
  {author} {\bibfnamefont {T.~F.}\ \bibnamefont {Heinz}}, \bibinfo {author}
  {\bibfnamefont {X.}~\bibnamefont {Marie}}, \bibinfo {author} {\bibfnamefont
  {T.}~\bibnamefont {Amand}}, \ and\ \bibinfo {author} {\bibfnamefont
  {B.}~\bibnamefont {Urbaszek}},\ }\href {\doibase
  10.1103/RevModPhys.90.021001} {\bibfield  {journal} {\bibinfo  {journal}
  {Rev. Mod. Phys.}\ }\textbf {\bibinfo {volume} {90}},\ \bibinfo {pages}
  {021001} (\bibinfo {year} {2018})}\BibitemShut {NoStop}%
\bibitem [{\citenamefont {Korm\'anyos}\ \emph {et~al.}(2015)\citenamefont
  {Korm\'anyos}, \citenamefont {Burkard}, \citenamefont {Gmitra}, \citenamefont
  {Fabian}, \citenamefont {Z\'olyomi}, \citenamefont {Drummond},\ and\
  \citenamefont {Fal'ko}}]{Kormanyos2015}%
  \BibitemOpen
  \bibfield  {author} {\bibinfo {author} {\bibfnamefont {A.}~\bibnamefont
  {Korm\'anyos}}, \bibinfo {author} {\bibfnamefont {G.}~\bibnamefont
  {Burkard}}, \bibinfo {author} {\bibfnamefont {M.}~\bibnamefont {Gmitra}},
  \bibinfo {author} {\bibfnamefont {J.}~\bibnamefont {Fabian}}, \bibinfo
  {author} {\bibfnamefont {V.}~\bibnamefont {Z\'olyomi}}, \bibinfo {author}
  {\bibfnamefont {N.~D.}\ \bibnamefont {Drummond}}, \ and\ \bibinfo {author}
  {\bibfnamefont {V.}~\bibnamefont {Fal'ko}},\ }\href {\doibase
  10.1088/2053-1583/2/2/022001} {\bibfield  {journal} {\bibinfo  {journal} {2D
  Materials}\ }\textbf {\bibinfo {volume} {2}},\ \bibinfo {pages} {022001}
  (\bibinfo {year} {2015})}\BibitemShut {NoStop}%
\bibitem [{\citenamefont {Molas}\ \emph
  {et~al.}(2017{\natexlab{a}})\citenamefont {Molas}, \citenamefont {Faugeras},
  \citenamefont {Slobodeniuk}, \citenamefont {Nogajewski}, \citenamefont
  {Bartos}, \citenamefont {Basko},\ and\ \citenamefont {Potemski}}]{Molas2017}%
  \BibitemOpen
  \bibfield  {author} {\bibinfo {author} {\bibfnamefont {M.~R.}\ \bibnamefont
  {Molas}}, \bibinfo {author} {\bibfnamefont {C.}~\bibnamefont {Faugeras}},
  \bibinfo {author} {\bibfnamefont {A.~O.}\ \bibnamefont {Slobodeniuk}},
  \bibinfo {author} {\bibfnamefont {K.}~\bibnamefont {Nogajewski}}, \bibinfo
  {author} {\bibfnamefont {M.}~\bibnamefont {Bartos}}, \bibinfo {author}
  {\bibfnamefont {D.~M.}\ \bibnamefont {Basko}}, \ and\ \bibinfo {author}
  {\bibfnamefont {M.}~\bibnamefont {Potemski}},\ }\href {\doibase
  10.1088/2053-1583/aa5521} {\bibfield  {journal} {\bibinfo  {journal} {2D
  Materials}\ }\textbf {\bibinfo {volume} {4}},\ \bibinfo {pages} {021003}
  (\bibinfo {year} {2017}{\natexlab{a}})}\BibitemShut {NoStop}%
\bibitem [{\citenamefont {Robert}\ \emph
  {et~al.}(2020{\natexlab{a}})\citenamefont {Robert}, \citenamefont {Han},
  \citenamefont {Kapuscinski}, \citenamefont {Delhomme}, \citenamefont
  {Faugeras}, \citenamefont {Amand}, \citenamefont {Molas}, \citenamefont
  {Bartos}, \citenamefont {Watanabe}, \citenamefont {Taniguchi}, \citenamefont
  {Urbaszek}, \citenamefont {Potemski},\ and\ \citenamefont
  {Marie}}]{Robert2020}%
  \BibitemOpen
  \bibfield  {author} {\bibinfo {author} {\bibfnamefont {C.}~\bibnamefont
  {Robert}}, \bibinfo {author} {\bibfnamefont {B.}~\bibnamefont {Han}},
  \bibinfo {author} {\bibfnamefont {P.}~\bibnamefont {Kapuscinski}}, \bibinfo
  {author} {\bibfnamefont {A.}~\bibnamefont {Delhomme}}, \bibinfo {author}
  {\bibfnamefont {C.}~\bibnamefont {Faugeras}}, \bibinfo {author}
  {\bibfnamefont {T.}~\bibnamefont {Amand}}, \bibinfo {author} {\bibfnamefont
  {M.~R.}\ \bibnamefont {Molas}}, \bibinfo {author} {\bibfnamefont
  {M.}~\bibnamefont {Bartos}}, \bibinfo {author} {\bibfnamefont
  {K.}~\bibnamefont {Watanabe}}, \bibinfo {author} {\bibfnamefont
  {T.}~\bibnamefont {Taniguchi}}, \bibinfo {author} {\bibfnamefont
  {B.}~\bibnamefont {Urbaszek}}, \bibinfo {author} {\bibfnamefont
  {M.}~\bibnamefont {Potemski}}, \ and\ \bibinfo {author} {\bibfnamefont
  {X.}~\bibnamefont {Marie}},\ }\href {\doibase 10.1038/s41467-020-17608-4}
  {\bibfield  {journal} {\bibinfo  {journal} {Nature Communications}\ }\textbf
  {\bibinfo {volume} {11}},\ \bibinfo {pages} {4037} (\bibinfo {year}
  {2020}{\natexlab{a}})}\BibitemShut {NoStop}%
\bibitem [{\citenamefont {Zhang}\ \emph {et~al.}(2017)\citenamefont {Zhang},
  \citenamefont {Cao}, \citenamefont {Lu}, \citenamefont {Lin}, \citenamefont
  {Zhang}, \citenamefont {Wang}, \citenamefont {Li}, \citenamefont {Hone},
  \citenamefont {Robinson}, \citenamefont {Smirnov}, \citenamefont {Louie},\
  and\ \citenamefont {Heinz}}]{Zhang2017}%
  \BibitemOpen
  \bibfield  {author} {\bibinfo {author} {\bibfnamefont {X.-X.}\ \bibnamefont
  {Zhang}}, \bibinfo {author} {\bibfnamefont {T.}~\bibnamefont {Cao}}, \bibinfo
  {author} {\bibfnamefont {Z.}~\bibnamefont {Lu}}, \bibinfo {author}
  {\bibfnamefont {Y.-C.}\ \bibnamefont {Lin}}, \bibinfo {author} {\bibfnamefont
  {F.}~\bibnamefont {Zhang}}, \bibinfo {author} {\bibfnamefont
  {Y.}~\bibnamefont {Wang}}, \bibinfo {author} {\bibfnamefont {Z.}~\bibnamefont
  {Li}}, \bibinfo {author} {\bibfnamefont {J.~C.}\ \bibnamefont {Hone}},
  \bibinfo {author} {\bibfnamefont {J.~A.}\ \bibnamefont {Robinson}}, \bibinfo
  {author} {\bibfnamefont {D.}~\bibnamefont {Smirnov}}, \bibinfo {author}
  {\bibfnamefont {S.~G.}\ \bibnamefont {Louie}}, \ and\ \bibinfo {author}
  {\bibfnamefont {T.~F.}\ \bibnamefont {Heinz}},\ }\href {\doibase
  10.1038/nnano.2017.105} {\bibfield  {journal} {\bibinfo  {journal} {Nature
  Nanotechnology}\ }\textbf {\bibinfo {volume} {12}},\ \bibinfo {pages} {883}
  (\bibinfo {year} {2017})}\BibitemShut {NoStop}%
\bibitem [{\citenamefont {Slobodeniuk}\ and\ \citenamefont
  {Basko}(2016)}]{Slobodeniuk2016}%
  \BibitemOpen
  \bibfield  {author} {\bibinfo {author} {\bibfnamefont {A.~O.}\ \bibnamefont
  {Slobodeniuk}}\ and\ \bibinfo {author} {\bibfnamefont {D.~M.}\ \bibnamefont
  {Basko}},\ }\href {\doibase 10.1088/2053-1583/3/3/035009} {\bibfield
  {journal} {\bibinfo  {journal} {2D Materials}\ }\textbf {\bibinfo {volume}
  {3}},\ \bibinfo {pages} {035009} (\bibinfo {year} {2016})}\BibitemShut
  {NoStop}%
\bibitem [{\citenamefont {Molas}\ \emph
  {et~al.}(2019{\natexlab{a}})\citenamefont {Molas}, \citenamefont
  {Slobodeniuk}, \citenamefont {Kazimierczuk}, \citenamefont {Nogajewski},
  \citenamefont {Bartos}, \citenamefont {Kapu\ifmmode \acute{s}\else
  \'{s}\fi{}ci\ifmmode~\acute{n}\else \'{n}\fi{}ski}, \citenamefont
  {Oreszczuk}, \citenamefont {Watanabe}, \citenamefont {Taniguchi},
  \citenamefont {Faugeras}, \citenamefont {Kossacki}, \citenamefont {Basko},\
  and\ \citenamefont {Potemski}}]{Molas2019Dark}%
  \BibitemOpen
  \bibfield  {author} {\bibinfo {author} {\bibfnamefont {M.~R.}\ \bibnamefont
  {Molas}}, \bibinfo {author} {\bibfnamefont {A.~O.}\ \bibnamefont
  {Slobodeniuk}}, \bibinfo {author} {\bibfnamefont {T.}~\bibnamefont
  {Kazimierczuk}}, \bibinfo {author} {\bibfnamefont {K.}~\bibnamefont
  {Nogajewski}}, \bibinfo {author} {\bibfnamefont {M.}~\bibnamefont {Bartos}},
  \bibinfo {author} {\bibfnamefont {P.}~\bibnamefont {Kapu\ifmmode
  \acute{s}\else \'{s}\fi{}ci\ifmmode~\acute{n}\else \'{n}\fi{}ski}}, \bibinfo
  {author} {\bibfnamefont {K.}~\bibnamefont {Oreszczuk}}, \bibinfo {author}
  {\bibfnamefont {K.}~\bibnamefont {Watanabe}}, \bibinfo {author}
  {\bibfnamefont {T.}~\bibnamefont {Taniguchi}}, \bibinfo {author}
  {\bibfnamefont {C.}~\bibnamefont {Faugeras}}, \bibinfo {author}
  {\bibfnamefont {P.}~\bibnamefont {Kossacki}}, \bibinfo {author}
  {\bibfnamefont {D.~M.}\ \bibnamefont {Basko}}, \ and\ \bibinfo {author}
  {\bibfnamefont {M.}~\bibnamefont {Potemski}},\ }\href {\doibase
  10.1103/PhysRevLett.123.096803} {\bibfield  {journal} {\bibinfo  {journal}
  {Phys. Rev. Lett.}\ }\textbf {\bibinfo {volume} {123}},\ \bibinfo {pages}
  {096803} (\bibinfo {year} {2019}{\natexlab{a}})}\BibitemShut {NoStop}%
\bibitem [{\citenamefont {Lu}\ \emph {et~al.}(2019)\citenamefont {Lu},
  \citenamefont {Rhodes}, \citenamefont {Li}, \citenamefont {Tuan},
  \citenamefont {Jiang}, \citenamefont {Ludwig}, \citenamefont {Jiang},
  \citenamefont {Lian}, \citenamefont {Shi}, \citenamefont {Hone},
  \citenamefont {Dery},\ and\ \citenamefont {Smirnov}}]{Lu2019}%
  \BibitemOpen
  \bibfield  {author} {\bibinfo {author} {\bibfnamefont {Z.}~\bibnamefont
  {Lu}}, \bibinfo {author} {\bibfnamefont {D.}~\bibnamefont {Rhodes}}, \bibinfo
  {author} {\bibfnamefont {Z.}~\bibnamefont {Li}}, \bibinfo {author}
  {\bibfnamefont {D.~V.}\ \bibnamefont {Tuan}}, \bibinfo {author}
  {\bibfnamefont {Y.}~\bibnamefont {Jiang}}, \bibinfo {author} {\bibfnamefont
  {J.}~\bibnamefont {Ludwig}}, \bibinfo {author} {\bibfnamefont
  {Z.}~\bibnamefont {Jiang}}, \bibinfo {author} {\bibfnamefont
  {Z.}~\bibnamefont {Lian}}, \bibinfo {author} {\bibfnamefont {S.-F.}\
  \bibnamefont {Shi}}, \bibinfo {author} {\bibfnamefont {J.}~\bibnamefont
  {Hone}}, \bibinfo {author} {\bibfnamefont {H.}~\bibnamefont {Dery}}, \ and\
  \bibinfo {author} {\bibfnamefont {D.}~\bibnamefont {Smirnov}},\ }\href
  {\doibase 10.1088/2053-1583/ab5614} {\bibfield  {journal} {\bibinfo
  {journal} {2D Materials}\ }\textbf {\bibinfo {volume} {7}},\ \bibinfo {pages}
  {015017} (\bibinfo {year} {2019})}\BibitemShut {NoStop}%
\bibitem [{\citenamefont {Zinkiewicz}\ \emph {et~al.}(2020)\citenamefont
  {Zinkiewicz}, \citenamefont {Slobodeniuk}, \citenamefont {Kazimierczuk},
  \citenamefont {Kapu{\'{s}}ci{\'{n}}ski}, \citenamefont {Oreszczuk},
  \citenamefont {Grzeszczyk}, \citenamefont {Bartos}, \citenamefont
  {Nogajewski}, \citenamefont {Watanabe}, \citenamefont {Taniguchi},
  \citenamefont {Faugeras}, \citenamefont {Kossacki}, \citenamefont {Potemski},
  \citenamefont {Babi{\'{n}}ski},\ and\ \citenamefont
  {Molas}}]{Zinkiewicz2020}%
  \BibitemOpen
  \bibfield  {author} {\bibinfo {author} {\bibfnamefont {M.}~\bibnamefont
  {Zinkiewicz}}, \bibinfo {author} {\bibfnamefont {A.~O.}\ \bibnamefont
  {Slobodeniuk}}, \bibinfo {author} {\bibfnamefont {T.}~\bibnamefont
  {Kazimierczuk}}, \bibinfo {author} {\bibfnamefont {P.}~\bibnamefont
  {Kapu{\'{s}}ci{\'{n}}ski}}, \bibinfo {author} {\bibfnamefont
  {K.}~\bibnamefont {Oreszczuk}}, \bibinfo {author} {\bibfnamefont
  {M.}~\bibnamefont {Grzeszczyk}}, \bibinfo {author} {\bibfnamefont
  {M.}~\bibnamefont {Bartos}}, \bibinfo {author} {\bibfnamefont
  {K.}~\bibnamefont {Nogajewski}}, \bibinfo {author} {\bibfnamefont
  {K.}~\bibnamefont {Watanabe}}, \bibinfo {author} {\bibfnamefont
  {T.}~\bibnamefont {Taniguchi}}, \bibinfo {author} {\bibfnamefont
  {C.}~\bibnamefont {Faugeras}}, \bibinfo {author} {\bibfnamefont
  {P.}~\bibnamefont {Kossacki}}, \bibinfo {author} {\bibfnamefont
  {M.}~\bibnamefont {Potemski}}, \bibinfo {author} {\bibfnamefont
  {A.}~\bibnamefont {Babi{\'{n}}ski}}, \ and\ \bibinfo {author} {\bibfnamefont
  {M.~R.}\ \bibnamefont {Molas}},\ }\href {\doibase 10.1039/d0nr04243a}
  {\bibfield  {journal} {\bibinfo  {journal} {Nanoscale}\ }\textbf {\bibinfo
  {volume} {12}},\ \bibinfo {pages} {18153} (\bibinfo {year}
  {2020})}\BibitemShut {NoStop}%
\bibitem [{\citenamefont {Vaclavkova}\ \emph {et~al.}(2018)\citenamefont
  {Vaclavkova}, \citenamefont {Wyzula}, \citenamefont {Nogajewski},
  \citenamefont {Bartos}, \citenamefont {Slobodeniuk}, \citenamefont
  {Faugeras}, \citenamefont {Potemski},\ and\ \citenamefont
  {Molas}}]{Vaclavkova2018}%
  \BibitemOpen
  \bibfield  {author} {\bibinfo {author} {\bibfnamefont {D.}~\bibnamefont
  {Vaclavkova}}, \bibinfo {author} {\bibfnamefont {J.}~\bibnamefont {Wyzula}},
  \bibinfo {author} {\bibfnamefont {K.}~\bibnamefont {Nogajewski}}, \bibinfo
  {author} {\bibfnamefont {M.}~\bibnamefont {Bartos}}, \bibinfo {author}
  {\bibfnamefont {A.~O.}\ \bibnamefont {Slobodeniuk}}, \bibinfo {author}
  {\bibfnamefont {C.}~\bibnamefont {Faugeras}}, \bibinfo {author}
  {\bibfnamefont {M.}~\bibnamefont {Potemski}}, \ and\ \bibinfo {author}
  {\bibfnamefont {M.~R.}\ \bibnamefont {Molas}},\ }\href {\doibase
  10.1088/1361-6528/aac65c} {\bibfield  {journal} {\bibinfo  {journal}
  {Nanotechnology}\ }\textbf {\bibinfo {volume} {29}},\ \bibinfo {pages}
  {325705} (\bibinfo {year} {2018})}\BibitemShut {NoStop}%
\bibitem [{\citenamefont {Nagler}\ \emph {et~al.}(2018)\citenamefont {Nagler},
  \citenamefont {Ballottin}, \citenamefont {Mitioglu}, \citenamefont {Durnev},
  \citenamefont {Taniguchi}, \citenamefont {Watanabe}, \citenamefont
  {Chernikov}, \citenamefont {Sch\"uller}, \citenamefont {Glazov},
  \citenamefont {Christianen},\ and\ \citenamefont {Korn}}]{nagler2018}%
  \BibitemOpen
  \bibfield  {author} {\bibinfo {author} {\bibfnamefont {P.}~\bibnamefont
  {Nagler}}, \bibinfo {author} {\bibfnamefont {M.~V.}\ \bibnamefont
  {Ballottin}}, \bibinfo {author} {\bibfnamefont {A.~A.}\ \bibnamefont
  {Mitioglu}}, \bibinfo {author} {\bibfnamefont {M.~V.}\ \bibnamefont
  {Durnev}}, \bibinfo {author} {\bibfnamefont {T.}~\bibnamefont {Taniguchi}},
  \bibinfo {author} {\bibfnamefont {K.}~\bibnamefont {Watanabe}}, \bibinfo
  {author} {\bibfnamefont {A.}~\bibnamefont {Chernikov}}, \bibinfo {author}
  {\bibfnamefont {C.}~\bibnamefont {Sch\"uller}}, \bibinfo {author}
  {\bibfnamefont {M.~M.}\ \bibnamefont {Glazov}}, \bibinfo {author}
  {\bibfnamefont {P.~C.~M.}\ \bibnamefont {Christianen}}, \ and\ \bibinfo
  {author} {\bibfnamefont {T.}~\bibnamefont {Korn}},\ }\href {\doibase
  10.1103/PhysRevLett.121.057402} {\bibfield  {journal} {\bibinfo  {journal}
  {Phys. Rev. Lett.}\ }\textbf {\bibinfo {volume} {121}},\ \bibinfo {pages}
  {057402} (\bibinfo {year} {2018})}\BibitemShut {NoStop}%
\bibitem [{\citenamefont {Jadczak}\ \emph {et~al.}(2019)\citenamefont
  {Jadczak}, \citenamefont {Bryja}, \citenamefont {Kutrowska-Girzycka},
  \citenamefont {Kapuscinski}, \citenamefont {Bieniek}, \citenamefont {Huang},\
  and\ \citenamefont {Hawrylak}}]{Jadczak2019}%
  \BibitemOpen
  \bibfield  {author} {\bibinfo {author} {\bibfnamefont {J.}~\bibnamefont
  {Jadczak}}, \bibinfo {author} {\bibfnamefont {L.}~\bibnamefont {Bryja}},
  \bibinfo {author} {\bibfnamefont {J.}~\bibnamefont {Kutrowska-Girzycka}},
  \bibinfo {author} {\bibfnamefont {P.}~\bibnamefont {Kapuscinski}}, \bibinfo
  {author} {\bibfnamefont {M.}~\bibnamefont {Bieniek}}, \bibinfo {author}
  {\bibfnamefont {Y.-S.}\ \bibnamefont {Huang}}, \ and\ \bibinfo {author}
  {\bibfnamefont {P.}~\bibnamefont {Hawrylak}},\ }\href {\doibase
  10.1038/s41467-018-07994-1} {\bibfield  {journal} {\bibinfo  {journal}
  {Nature Communications}\ }\textbf {\bibinfo {volume} {10}},\ \bibinfo {pages}
  {107} (\bibinfo {year} {2019})}\BibitemShut {NoStop}%
\bibitem [{\citenamefont {Paur}\ \emph {et~al.}(2019)\citenamefont {Paur},
  \citenamefont {Molina-Mendoza}, \citenamefont {Bratschitsch}, \citenamefont
  {Watanabe}, \citenamefont {Taniguchi},\ and\ \citenamefont
  {Mueller}}]{Paur2019}%
  \BibitemOpen
  \bibfield  {author} {\bibinfo {author} {\bibfnamefont {M.}~\bibnamefont
  {Paur}}, \bibinfo {author} {\bibfnamefont {A.~J.}\ \bibnamefont
  {Molina-Mendoza}}, \bibinfo {author} {\bibfnamefont {R.}~\bibnamefont
  {Bratschitsch}}, \bibinfo {author} {\bibfnamefont {K.}~\bibnamefont
  {Watanabe}}, \bibinfo {author} {\bibfnamefont {T.}~\bibnamefont {Taniguchi}},
  \ and\ \bibinfo {author} {\bibfnamefont {T.}~\bibnamefont {Mueller}},\ }\href
  {\doibase 10.1038/s41467-019-09781-y} {\bibfield  {journal} {\bibinfo
  {journal} {Nature Communications}\ }\textbf {\bibinfo {volume} {10}},\
  \bibinfo {pages} {1709} (\bibinfo {year} {2019})}\BibitemShut {NoStop}%
\bibitem [{\citenamefont {He}\ \emph {et~al.}(2020{\natexlab{a}})\citenamefont
  {He}, \citenamefont {Rivera}, \citenamefont {Van~Tuan}, \citenamefont
  {Wilson}, \citenamefont {Yang}, \citenamefont {Taniguchi}, \citenamefont
  {Watanabe}, \citenamefont {Yan}, \citenamefont {Mandrus}, \citenamefont {Yu},
  \citenamefont {Dery}, \citenamefont {Yao},\ and\ \citenamefont
  {Xu}}]{he2020valley}%
  \BibitemOpen
  \bibfield  {author} {\bibinfo {author} {\bibfnamefont {M.}~\bibnamefont
  {He}}, \bibinfo {author} {\bibfnamefont {P.}~\bibnamefont {Rivera}}, \bibinfo
  {author} {\bibfnamefont {D.}~\bibnamefont {Van~Tuan}}, \bibinfo {author}
  {\bibfnamefont {N.~P.}\ \bibnamefont {Wilson}}, \bibinfo {author}
  {\bibfnamefont {M.}~\bibnamefont {Yang}}, \bibinfo {author} {\bibfnamefont
  {T.}~\bibnamefont {Taniguchi}}, \bibinfo {author} {\bibfnamefont
  {K.}~\bibnamefont {Watanabe}}, \bibinfo {author} {\bibfnamefont
  {J.}~\bibnamefont {Yan}}, \bibinfo {author} {\bibfnamefont {D.~G.}\
  \bibnamefont {Mandrus}}, \bibinfo {author} {\bibfnamefont {H.}~\bibnamefont
  {Yu}}, \bibinfo {author} {\bibfnamefont {H.}~\bibnamefont {Dery}}, \bibinfo
  {author} {\bibfnamefont {W.}~\bibnamefont {Yao}}, \ and\ \bibinfo {author}
  {\bibfnamefont {X.}~\bibnamefont {Xu}},\ }\href {\doibase
  10.1038/s41467-020-14472-0} {\bibfield  {journal} {\bibinfo  {journal}
  {Nature Communications}\ }\textbf {\bibinfo {volume} {11}},\ \bibinfo {pages}
  {618} (\bibinfo {year} {2020}{\natexlab{a}})}\BibitemShut {NoStop}%
\bibitem [{\citenamefont {Liu}\ \emph {et~al.}(2020)\citenamefont {Liu},
  \citenamefont {van Baren}, \citenamefont {Liang}, \citenamefont {Taniguchi},
  \citenamefont {Watanabe}, \citenamefont {Gabor}, \citenamefont {Chang},\ and\
  \citenamefont {Lui}}]{Liu2020}%
  \BibitemOpen
  \bibfield  {author} {\bibinfo {author} {\bibfnamefont {E.}~\bibnamefont
  {Liu}}, \bibinfo {author} {\bibfnamefont {J.}~\bibnamefont {van Baren}},
  \bibinfo {author} {\bibfnamefont {C.-T.}\ \bibnamefont {Liang}}, \bibinfo
  {author} {\bibfnamefont {T.}~\bibnamefont {Taniguchi}}, \bibinfo {author}
  {\bibfnamefont {K.}~\bibnamefont {Watanabe}}, \bibinfo {author}
  {\bibfnamefont {N.~M.}\ \bibnamefont {Gabor}}, \bibinfo {author}
  {\bibfnamefont {Y.-C.}\ \bibnamefont {Chang}}, \ and\ \bibinfo {author}
  {\bibfnamefont {C.~H.}\ \bibnamefont {Lui}},\ }\href {\doibase
  10.1103/physrevlett.124.196802} {\bibfield  {journal} {\bibinfo  {journal}
  {Physical Review Letters}\ }\textbf {\bibinfo {volume} {124}},\ \bibinfo
  {pages} {196802} (\bibinfo {year} {2020})}\BibitemShut {NoStop}%
\bibitem [{\citenamefont {Danovich}\ \emph {et~al.}(2016)\citenamefont
  {Danovich}, \citenamefont {Z{\'{o}}lyomi}, \citenamefont {Fal'ko},\ and\
  \citenamefont {Aleiner}}]{Danovich_2016}%
  \BibitemOpen
  \bibfield  {author} {\bibinfo {author} {\bibfnamefont {M.}~\bibnamefont
  {Danovich}}, \bibinfo {author} {\bibfnamefont {V.}~\bibnamefont
  {Z{\'{o}}lyomi}}, \bibinfo {author} {\bibfnamefont {V.~I.}\ \bibnamefont
  {Fal'ko}}, \ and\ \bibinfo {author} {\bibfnamefont {I.~L.}\ \bibnamefont
  {Aleiner}},\ }\href {\doibase 10.1088/2053-1583/3/3/035011} {\bibfield
  {journal} {\bibinfo  {journal} {2D Materials}\ }\textbf {\bibinfo {volume}
  {3}},\ \bibinfo {pages} {035011} (\bibinfo {year} {2016})}\BibitemShut
  {NoStop}%
\bibitem [{\citenamefont {Molas}\ \emph
  {et~al.}(2017{\natexlab{b}})\citenamefont {Molas}, \citenamefont
  {Nogajewski}, \citenamefont {Slobodeniuk}, \citenamefont {Binder},
  \citenamefont {Bartos},\ and\ \citenamefont {Potemski}}]{Molas2017nano}%
  \BibitemOpen
  \bibfield  {author} {\bibinfo {author} {\bibfnamefont {M.~R.}\ \bibnamefont
  {Molas}}, \bibinfo {author} {\bibfnamefont {K.}~\bibnamefont {Nogajewski}},
  \bibinfo {author} {\bibfnamefont {A.~O.}\ \bibnamefont {Slobodeniuk}},
  \bibinfo {author} {\bibfnamefont {J.}~\bibnamefont {Binder}}, \bibinfo
  {author} {\bibfnamefont {M.}~\bibnamefont {Bartos}}, \ and\ \bibinfo {author}
  {\bibfnamefont {M.}~\bibnamefont {Potemski}},\ }\href {\doibase
  10.1039/c7nr04672c} {\bibfield  {journal} {\bibinfo  {journal} {Nanoscale}\
  }\textbf {\bibinfo {volume} {9}},\ \bibinfo {pages} {13128} (\bibinfo {year}
  {2017}{\natexlab{b}})}\BibitemShut {NoStop}%
\bibitem [{\citenamefont {Li}\ \emph {et~al.}(2019{\natexlab{a}})\citenamefont
  {Li}, \citenamefont {Wang}, \citenamefont {Jin}, \citenamefont {Lu},
  \citenamefont {Lian}, \citenamefont {Meng}, \citenamefont {Blei},
  \citenamefont {Gao}, \citenamefont {Taniguchi}, \citenamefont {Watanabe},
  \citenamefont {Ren}, \citenamefont {Cao}, \citenamefont {Tongay},
  \citenamefont {Smirnov}, \citenamefont {Zhang},\ and\ \citenamefont
  {Shi}}]{Li2019Momentum}%
  \BibitemOpen
  \bibfield  {author} {\bibinfo {author} {\bibfnamefont {Z.}~\bibnamefont
  {Li}}, \bibinfo {author} {\bibfnamefont {T.}~\bibnamefont {Wang}}, \bibinfo
  {author} {\bibfnamefont {C.}~\bibnamefont {Jin}}, \bibinfo {author}
  {\bibfnamefont {Z.}~\bibnamefont {Lu}}, \bibinfo {author} {\bibfnamefont
  {Z.}~\bibnamefont {Lian}}, \bibinfo {author} {\bibfnamefont {Y.}~\bibnamefont
  {Meng}}, \bibinfo {author} {\bibfnamefont {M.}~\bibnamefont {Blei}}, \bibinfo
  {author} {\bibfnamefont {M.}~\bibnamefont {Gao}}, \bibinfo {author}
  {\bibfnamefont {T.}~\bibnamefont {Taniguchi}}, \bibinfo {author}
  {\bibfnamefont {K.}~\bibnamefont {Watanabe}}, \bibinfo {author}
  {\bibfnamefont {T.}~\bibnamefont {Ren}}, \bibinfo {author} {\bibfnamefont
  {T.}~\bibnamefont {Cao}}, \bibinfo {author} {\bibfnamefont {S.}~\bibnamefont
  {Tongay}}, \bibinfo {author} {\bibfnamefont {D.}~\bibnamefont {Smirnov}},
  \bibinfo {author} {\bibfnamefont {L.}~\bibnamefont {Zhang}}, \ and\ \bibinfo
  {author} {\bibfnamefont {S.-F.}\ \bibnamefont {Shi}},\ }\href {\doibase
  10.1021/acsnano.9b06682} {\bibfield  {journal} {\bibinfo  {journal} {ACS
  Nano}\ }\textbf {\bibinfo {volume} {13}},\ \bibinfo {pages} {14107} (\bibinfo
  {year} {2019}{\natexlab{a}})}\BibitemShut {NoStop}%
\bibitem [{\citenamefont {Li}\ \emph {et~al.}(2019{\natexlab{b}})\citenamefont
  {Li}, \citenamefont {Wang}, \citenamefont {Jin}, \citenamefont {Lu},
  \citenamefont {Lian}, \citenamefont {Meng}, \citenamefont {Blei},
  \citenamefont {Gao}, \citenamefont {Taniguchi}, \citenamefont {Watanabe},
  \citenamefont {Ren}, \citenamefont {Tongay}, \citenamefont {Yang},
  \citenamefont {Smirnov}, \citenamefont {Cao},\ and\ \citenamefont
  {Shi}}]{Li2019Replica}%
  \BibitemOpen
  \bibfield  {author} {\bibinfo {author} {\bibfnamefont {Z.}~\bibnamefont
  {Li}}, \bibinfo {author} {\bibfnamefont {T.}~\bibnamefont {Wang}}, \bibinfo
  {author} {\bibfnamefont {C.}~\bibnamefont {Jin}}, \bibinfo {author}
  {\bibfnamefont {Z.}~\bibnamefont {Lu}}, \bibinfo {author} {\bibfnamefont
  {Z.}~\bibnamefont {Lian}}, \bibinfo {author} {\bibfnamefont {Y.}~\bibnamefont
  {Meng}}, \bibinfo {author} {\bibfnamefont {M.}~\bibnamefont {Blei}}, \bibinfo
  {author} {\bibfnamefont {S.}~\bibnamefont {Gao}}, \bibinfo {author}
  {\bibfnamefont {T.}~\bibnamefont {Taniguchi}}, \bibinfo {author}
  {\bibfnamefont {K.}~\bibnamefont {Watanabe}}, \bibinfo {author}
  {\bibfnamefont {T.}~\bibnamefont {Ren}}, \bibinfo {author} {\bibfnamefont
  {S.}~\bibnamefont {Tongay}}, \bibinfo {author} {\bibfnamefont
  {L.}~\bibnamefont {Yang}}, \bibinfo {author} {\bibfnamefont {D.}~\bibnamefont
  {Smirnov}}, \bibinfo {author} {\bibfnamefont {T.}~\bibnamefont {Cao}}, \ and\
  \bibinfo {author} {\bibfnamefont {S.-F.}\ \bibnamefont {Shi}},\ }\href
  {\doibase 10.1038/s41467-019-10477-6} {\bibfield  {journal} {\bibinfo
  {journal} {Nature Communications}\ }\textbf {\bibinfo {volume} {10}},\
  \bibinfo {pages} {2469} (\bibinfo {year} {2019}{\natexlab{b}})}\BibitemShut
  {NoStop}%
\bibitem [{\citenamefont {Liu}\ \emph {et~al.}(2019{\natexlab{a}})\citenamefont
  {Liu}, \citenamefont {van Baren}, \citenamefont {Taniguchi}, \citenamefont
  {Watanabe}, \citenamefont {Chang},\ and\ \citenamefont
  {Lui}}]{Liu2019Replica}%
  \BibitemOpen
  \bibfield  {author} {\bibinfo {author} {\bibfnamefont {E.}~\bibnamefont
  {Liu}}, \bibinfo {author} {\bibfnamefont {J.}~\bibnamefont {van Baren}},
  \bibinfo {author} {\bibfnamefont {T.}~\bibnamefont {Taniguchi}}, \bibinfo
  {author} {\bibfnamefont {K.}~\bibnamefont {Watanabe}}, \bibinfo {author}
  {\bibfnamefont {Y.-C.}\ \bibnamefont {Chang}}, \ and\ \bibinfo {author}
  {\bibfnamefont {C.~H.}\ \bibnamefont {Lui}},\ }\href {\doibase
  10.1103/PhysRevResearch.1.032007} {\bibfield  {journal} {\bibinfo  {journal}
  {Phys. Rev. Research}\ }\textbf {\bibinfo {volume} {1}},\ \bibinfo {pages}
  {032007} (\bibinfo {year} {2019}{\natexlab{a}})}\BibitemShut {NoStop}%
\bibitem [{\citenamefont {Robert}\ \emph
  {et~al.}(2020{\natexlab{b}})\citenamefont {Robert}, \citenamefont {Dery},
  \citenamefont {Ren}, \citenamefont {van Tuan}, \citenamefont {Courtade},
  \citenamefont {Yang}, \citenamefont {Urbaszek}, \citenamefont {Lagarde},
  \citenamefont {Watanabe}, \citenamefont {Taniguchi}, \citenamefont {Amand},\
  and\ \citenamefont {Marie}}]{Robert2020gfactor}%
  \BibitemOpen
  \bibfield  {author} {\bibinfo {author} {\bibfnamefont {C.}~\bibnamefont
  {Robert}}, \bibinfo {author} {\bibfnamefont {H.}~\bibnamefont {Dery}},
  \bibinfo {author} {\bibfnamefont {L.}~\bibnamefont {Ren}}, \bibinfo {author}
  {\bibfnamefont {D.}~\bibnamefont {van Tuan}}, \bibinfo {author}
  {\bibfnamefont {E.}~\bibnamefont {Courtade}}, \bibinfo {author}
  {\bibfnamefont {M.}~\bibnamefont {Yang}}, \bibinfo {author} {\bibfnamefont
  {B.}~\bibnamefont {Urbaszek}}, \bibinfo {author} {\bibfnamefont
  {D.}~\bibnamefont {Lagarde}}, \bibinfo {author} {\bibfnamefont
  {K.}~\bibnamefont {Watanabe}}, \bibinfo {author} {\bibfnamefont
  {T.}~\bibnamefont {Taniguchi}}, \bibinfo {author} {\bibfnamefont
  {T.}~\bibnamefont {Amand}}, \ and\ \bibinfo {author} {\bibfnamefont
  {X.}~\bibnamefont {Marie}},\ }\href@noop {} {\enquote {\bibinfo {title}
  {Measurement of conduction and valence bands g-factors in a transition metal
  dichalcogenide monolayer},}\ } (\bibinfo {year} {2020}{\natexlab{b}}),\
  \Eprint {http://arxiv.org/abs/2008.07464} {arXiv:2008.07464
  [cond-mat.mtrl-sci]} \BibitemShut {NoStop}%
\bibitem [{\citenamefont {He}\ \emph {et~al.}(2020{\natexlab{b}})\citenamefont
  {He}, \citenamefont {Rivera}, \citenamefont {{Van Tuan}}, \citenamefont
  {Wilson}, \citenamefont {Yang}, \citenamefont {Taniguchi}, \citenamefont
  {Watanabe}, \citenamefont {Yan}, \citenamefont {Mandrus}, \citenamefont {Yu},
  \citenamefont {Dery}, \citenamefont {Yao},\ and\ \citenamefont
  {Xu}}]{He2020}%
  \BibitemOpen
  \bibfield  {author} {\bibinfo {author} {\bibfnamefont {M.}~\bibnamefont
  {He}}, \bibinfo {author} {\bibfnamefont {P.}~\bibnamefont {Rivera}}, \bibinfo
  {author} {\bibfnamefont {D.}~\bibnamefont {{Van Tuan}}}, \bibinfo {author}
  {\bibfnamefont {N.~P.}\ \bibnamefont {Wilson}}, \bibinfo {author}
  {\bibfnamefont {M.}~\bibnamefont {Yang}}, \bibinfo {author} {\bibfnamefont
  {T.}~\bibnamefont {Taniguchi}}, \bibinfo {author} {\bibfnamefont
  {K.}~\bibnamefont {Watanabe}}, \bibinfo {author} {\bibfnamefont
  {J.}~\bibnamefont {Yan}}, \bibinfo {author} {\bibfnamefont {D.~G.}\
  \bibnamefont {Mandrus}}, \bibinfo {author} {\bibfnamefont {H.}~\bibnamefont
  {Yu}}, \bibinfo {author} {\bibfnamefont {H.}~\bibnamefont {Dery}}, \bibinfo
  {author} {\bibfnamefont {W.}~\bibnamefont {Yao}}, \ and\ \bibinfo {author}
  {\bibfnamefont {X.}~\bibnamefont {Xu}},\ }\href {\doibase
  10.1038/s41467-020-14472-0} {\bibfield  {journal} {\bibinfo  {journal}
  {Nature Communications}\ }\textbf {\bibinfo {volume} {11}},\ \bibinfo {pages}
  {1} (\bibinfo {year} {2020}{\natexlab{b}})}\BibitemShut {NoStop}%
\bibitem [{\citenamefont {Singh}\ \emph {et~al.}(2016)\citenamefont {Singh},
  \citenamefont {Tran}, \citenamefont {Kolarczik}, \citenamefont {Seifert},
  \citenamefont {Wang}, \citenamefont {Hao}, \citenamefont {Pleskot},
  \citenamefont {Gabor}, \citenamefont {Helmrich}, \citenamefont {Owschimikow},
  \citenamefont {Woggon},\ and\ \citenamefont {Li}}]{Singh2016}%
  \BibitemOpen
  \bibfield  {author} {\bibinfo {author} {\bibfnamefont {A.}~\bibnamefont
  {Singh}}, \bibinfo {author} {\bibfnamefont {K.}~\bibnamefont {Tran}},
  \bibinfo {author} {\bibfnamefont {M.}~\bibnamefont {Kolarczik}}, \bibinfo
  {author} {\bibfnamefont {J.}~\bibnamefont {Seifert}}, \bibinfo {author}
  {\bibfnamefont {Y.}~\bibnamefont {Wang}}, \bibinfo {author} {\bibfnamefont
  {K.}~\bibnamefont {Hao}}, \bibinfo {author} {\bibfnamefont {D.}~\bibnamefont
  {Pleskot}}, \bibinfo {author} {\bibfnamefont {N.~M.}\ \bibnamefont {Gabor}},
  \bibinfo {author} {\bibfnamefont {S.}~\bibnamefont {Helmrich}}, \bibinfo
  {author} {\bibfnamefont {N.}~\bibnamefont {Owschimikow}}, \bibinfo {author}
  {\bibfnamefont {U.}~\bibnamefont {Woggon}}, \ and\ \bibinfo {author}
  {\bibfnamefont {X.}~\bibnamefont {Li}},\ }\href {\doibase
  10.1103/PhysRevLett.117.257402} {\bibfield  {journal} {\bibinfo  {journal}
  {Phys. Rev. Lett.}\ }\textbf {\bibinfo {volume} {117}},\ \bibinfo {pages}
  {257402} (\bibinfo {year} {2016})}\BibitemShut {NoStop}%
\bibitem [{\citenamefont {Li}\ \emph {et~al.}(2019{\natexlab{c}})\citenamefont
  {Li}, \citenamefont {Wang}, \citenamefont {Lu}, \citenamefont {Khatoniar},
  \citenamefont {Lian}, \citenamefont {Meng}, \citenamefont {Blei},
  \citenamefont {Taniguchi}, \citenamefont {Watanabe}, \citenamefont {McGill},
  \citenamefont {Tongay}, \citenamefont {Menon}, \citenamefont {Smirnov},\ and\
  \citenamefont {Shi}}]{Li2019Trion}%
  \BibitemOpen
  \bibfield  {author} {\bibinfo {author} {\bibfnamefont {Z.}~\bibnamefont
  {Li}}, \bibinfo {author} {\bibfnamefont {T.}~\bibnamefont {Wang}}, \bibinfo
  {author} {\bibfnamefont {Z.}~\bibnamefont {Lu}}, \bibinfo {author}
  {\bibfnamefont {M.}~\bibnamefont {Khatoniar}}, \bibinfo {author}
  {\bibfnamefont {Z.}~\bibnamefont {Lian}}, \bibinfo {author} {\bibfnamefont
  {Y.}~\bibnamefont {Meng}}, \bibinfo {author} {\bibfnamefont {M.}~\bibnamefont
  {Blei}}, \bibinfo {author} {\bibfnamefont {T.}~\bibnamefont {Taniguchi}},
  \bibinfo {author} {\bibfnamefont {K.}~\bibnamefont {Watanabe}}, \bibinfo
  {author} {\bibfnamefont {S.~A.}\ \bibnamefont {McGill}}, \bibinfo {author}
  {\bibfnamefont {S.}~\bibnamefont {Tongay}}, \bibinfo {author} {\bibfnamefont
  {V.~M.}\ \bibnamefont {Menon}}, \bibinfo {author} {\bibfnamefont
  {D.}~\bibnamefont {Smirnov}}, \ and\ \bibinfo {author} {\bibfnamefont
  {S.-F.}\ \bibnamefont {Shi}},\ }\href
  {https://doi.org/10.1021/acs.nanolett.9b02132} {\bibfield  {journal}
  {\bibinfo  {journal} {Nano Letters}\ }\textbf {\bibinfo {volume} {19}},\
  \bibinfo {pages} {6886} (\bibinfo {year} {2019}{\natexlab{c}})}\BibitemShut
  {NoStop}%
\bibitem [{\citenamefont {Liu}\ \emph {et~al.}(2019{\natexlab{b}})\citenamefont
  {Liu}, \citenamefont {van Baren}, \citenamefont {Lu}, \citenamefont
  {Altaiary}, \citenamefont {Taniguchi}, \citenamefont {Watanabe},
  \citenamefont {Smirnov},\ and\ \citenamefont {Lui}}]{Liu2019}%
  \BibitemOpen
  \bibfield  {author} {\bibinfo {author} {\bibfnamefont {E.}~\bibnamefont
  {Liu}}, \bibinfo {author} {\bibfnamefont {J.}~\bibnamefont {van Baren}},
  \bibinfo {author} {\bibfnamefont {Z.}~\bibnamefont {Lu}}, \bibinfo {author}
  {\bibfnamefont {M.~M.}\ \bibnamefont {Altaiary}}, \bibinfo {author}
  {\bibfnamefont {T.}~\bibnamefont {Taniguchi}}, \bibinfo {author}
  {\bibfnamefont {K.}~\bibnamefont {Watanabe}}, \bibinfo {author}
  {\bibfnamefont {D.}~\bibnamefont {Smirnov}}, \ and\ \bibinfo {author}
  {\bibfnamefont {C.~H.}\ \bibnamefont {Lui}},\ }\href {\doibase
  10.1103/PhysRevLett.123.027401} {\bibfield  {journal} {\bibinfo  {journal}
  {Phys. Rev. Lett.}\ }\textbf {\bibinfo {volume} {123}},\ \bibinfo {pages}
  {027401} (\bibinfo {year} {2019}{\natexlab{b}})}\BibitemShut {NoStop}%
\bibitem [{\citenamefont {Danovich}\ \emph {et~al.}(2017)\citenamefont
  {Danovich}, \citenamefont {Z{\'{o}}lyomi},\ and\ \citenamefont
  {Fal'ko}}]{Danovich2017}%
  \BibitemOpen
  \bibfield  {author} {\bibinfo {author} {\bibfnamefont {M.}~\bibnamefont
  {Danovich}}, \bibinfo {author} {\bibfnamefont {V.}~\bibnamefont
  {Z{\'{o}}lyomi}}, \ and\ \bibinfo {author} {\bibfnamefont {V.~I.}\
  \bibnamefont {Fal'ko}},\ }\href {\doibase 10.1038/srep45998} {\bibfield
  {journal} {\bibinfo  {journal} {Scientific Reports}\ }\textbf {\bibinfo
  {volume} {7}},\ \bibinfo {pages} {45998} (\bibinfo {year}
  {2017})}\BibitemShut {NoStop}%
\bibitem [{\citenamefont {Tu}\ \emph {et~al.}(2019)\citenamefont {Tu},
  \citenamefont {Borghardt}, \citenamefont {Grützmacher},\ and\ \citenamefont
  {Kardyna{\l}}}]{Tu2019}%
  \BibitemOpen
  \bibfield  {author} {\bibinfo {author} {\bibfnamefont {J.-S.}\ \bibnamefont
  {Tu}}, \bibinfo {author} {\bibfnamefont {S.}~\bibnamefont {Borghardt}},
  \bibinfo {author} {\bibfnamefont {D.}~\bibnamefont {Grützmacher}}, \ and\
  \bibinfo {author} {\bibfnamefont {B.~E.}\ \bibnamefont {Kardyna{\l}}},\
  }\href {\doibase 10.1088/1361-648x/ab2f56} {\bibfield  {journal} {\bibinfo
  {journal} {Journal of Physics: Condensed Matter}\ }\textbf {\bibinfo {volume}
  {31}},\ \bibinfo {pages} {415701} (\bibinfo {year} {2019})}\BibitemShut
  {NoStop}%
\bibitem [{\citenamefont {Barbone}\ \emph {et~al.}(2018)\citenamefont
  {Barbone}, \citenamefont {Montblanch}, \citenamefont {Kara}, \citenamefont
  {Palacios-Berraquero}, \citenamefont {Cadore}, \citenamefont {De~Fazio},
  \citenamefont {Pingault}, \citenamefont {Mostaani}, \citenamefont {Li},
  \citenamefont {Chen}, \citenamefont {Watanabe}, \citenamefont {Taniguchi},
  \citenamefont {Tongay}, \citenamefont {Wang}, \citenamefont {Ferrari},\ and\
  \citenamefont {Atat{\"u}re}}]{Barbone2018}%
  \BibitemOpen
  \bibfield  {author} {\bibinfo {author} {\bibfnamefont {M.}~\bibnamefont
  {Barbone}}, \bibinfo {author} {\bibfnamefont {A.~R.~P.}\ \bibnamefont
  {Montblanch}}, \bibinfo {author} {\bibfnamefont {D.~M.}\ \bibnamefont
  {Kara}}, \bibinfo {author} {\bibfnamefont {C.}~\bibnamefont
  {Palacios-Berraquero}}, \bibinfo {author} {\bibfnamefont {A.~R.}\
  \bibnamefont {Cadore}}, \bibinfo {author} {\bibfnamefont {D.}~\bibnamefont
  {De~Fazio}}, \bibinfo {author} {\bibfnamefont {B.}~\bibnamefont {Pingault}},
  \bibinfo {author} {\bibfnamefont {E.}~\bibnamefont {Mostaani}}, \bibinfo
  {author} {\bibfnamefont {H.}~\bibnamefont {Li}}, \bibinfo {author}
  {\bibfnamefont {B.}~\bibnamefont {Chen}}, \bibinfo {author} {\bibfnamefont
  {K.}~\bibnamefont {Watanabe}}, \bibinfo {author} {\bibfnamefont
  {T.}~\bibnamefont {Taniguchi}}, \bibinfo {author} {\bibfnamefont
  {S.}~\bibnamefont {Tongay}}, \bibinfo {author} {\bibfnamefont
  {G.}~\bibnamefont {Wang}}, \bibinfo {author} {\bibfnamefont {A.~C.}\
  \bibnamefont {Ferrari}}, \ and\ \bibinfo {author} {\bibfnamefont
  {M.}~\bibnamefont {Atat{\"u}re}},\ }\href {\doibase
  10.1038/s41467-018-05632-4} {\bibfield  {journal} {\bibinfo  {journal}
  {Nature Communications}\ }\textbf {\bibinfo {volume} {9}},\ \bibinfo {pages}
  {3721} (\bibinfo {year} {2018})}\BibitemShut {NoStop}%
\bibitem [{\citenamefont {Chen}\ \emph {et~al.}(2018)\citenamefont {Chen},
  \citenamefont {Goldstein}, \citenamefont {Taniguchi}, \citenamefont
  {Watanabe},\ and\ \citenamefont {Yan}}]{Chen2018}%
  \BibitemOpen
  \bibfield  {author} {\bibinfo {author} {\bibfnamefont {S.-Y.}\ \bibnamefont
  {Chen}}, \bibinfo {author} {\bibfnamefont {T.}~\bibnamefont {Goldstein}},
  \bibinfo {author} {\bibfnamefont {T.}~\bibnamefont {Taniguchi}}, \bibinfo
  {author} {\bibfnamefont {K.}~\bibnamefont {Watanabe}}, \ and\ \bibinfo
  {author} {\bibfnamefont {J.}~\bibnamefont {Yan}},\ }\href {\doibase
  10.1038/s41467-018-05558-x} {\bibfield  {journal} {\bibinfo  {journal}
  {Nature Communications}\ }\textbf {\bibinfo {volume} {9}},\ \bibinfo {pages}
  {3717} (\bibinfo {year} {2018})}\BibitemShut {NoStop}%
\bibitem [{\citenamefont {Li}\ \emph {et~al.}(2018)\citenamefont {Li},
  \citenamefont {Wang}, \citenamefont {Lu}, \citenamefont {Jin}, \citenamefont
  {Chen}, \citenamefont {Meng}, \citenamefont {Lian}, \citenamefont
  {Taniguchi}, \citenamefont {Watanabe}, \citenamefont {Zhang}, \citenamefont
  {Smirnov},\ and\ \citenamefont {Shi}}]{Li2018}%
  \BibitemOpen
  \bibfield  {author} {\bibinfo {author} {\bibfnamefont {Z.}~\bibnamefont
  {Li}}, \bibinfo {author} {\bibfnamefont {T.}~\bibnamefont {Wang}}, \bibinfo
  {author} {\bibfnamefont {Z.}~\bibnamefont {Lu}}, \bibinfo {author}
  {\bibfnamefont {C.}~\bibnamefont {Jin}}, \bibinfo {author} {\bibfnamefont
  {Y.}~\bibnamefont {Chen}}, \bibinfo {author} {\bibfnamefont {Y.}~\bibnamefont
  {Meng}}, \bibinfo {author} {\bibfnamefont {Z.}~\bibnamefont {Lian}}, \bibinfo
  {author} {\bibfnamefont {T.}~\bibnamefont {Taniguchi}}, \bibinfo {author}
  {\bibfnamefont {K.}~\bibnamefont {Watanabe}}, \bibinfo {author}
  {\bibfnamefont {S.}~\bibnamefont {Zhang}}, \bibinfo {author} {\bibfnamefont
  {D.}~\bibnamefont {Smirnov}}, \ and\ \bibinfo {author} {\bibfnamefont
  {S.-F.}\ \bibnamefont {Shi}},\ }\href {\doibase 10.1038/s41467-018-05863-5}
  {\bibfield  {journal} {\bibinfo  {journal} {Nature Communications}\ }\textbf
  {\bibinfo {volume} {9}},\ \bibinfo {pages} {3719} (\bibinfo {year}
  {2018})}\BibitemShut {NoStop}%
\bibitem [{\citenamefont {Stier}\ \emph {et~al.}(2016)\citenamefont {Stier},
  \citenamefont {McCreary}, \citenamefont {Jonker}, \citenamefont {Kono},\ and\
  \citenamefont {Crooker}}]{Stier2016}%
  \BibitemOpen
  \bibfield  {author} {\bibinfo {author} {\bibfnamefont {A.~V.}\ \bibnamefont
  {Stier}}, \bibinfo {author} {\bibfnamefont {K.~M.}\ \bibnamefont {McCreary}},
  \bibinfo {author} {\bibfnamefont {B.~T.}\ \bibnamefont {Jonker}}, \bibinfo
  {author} {\bibfnamefont {J.}~\bibnamefont {Kono}}, \ and\ \bibinfo {author}
  {\bibfnamefont {S.~A.}\ \bibnamefont {Crooker}},\ }\href {\doibase
  10.1038/ncomms10643} {\bibfield  {journal} {\bibinfo  {journal} {Nature
  Communications}\ }\textbf {\bibinfo {volume} {7}},\ \bibinfo {pages} {10643}
  (\bibinfo {year} {2016})}\BibitemShut {NoStop}%
\bibitem [{\citenamefont {Aivazian}\ \emph {et~al.}(2015)\citenamefont
  {Aivazian}, \citenamefont {Gong}, \citenamefont {Jones}, \citenamefont {Chu},
  \citenamefont {Yan}, \citenamefont {Mandrus}, \citenamefont {Zhang},
  \citenamefont {Cobden}, \citenamefont {Yao},\ and\ \citenamefont
  {Xu}}]{Aivazian2015}%
  \BibitemOpen
  \bibfield  {author} {\bibinfo {author} {\bibfnamefont {G.}~\bibnamefont
  {Aivazian}}, \bibinfo {author} {\bibfnamefont {Z.}~\bibnamefont {Gong}},
  \bibinfo {author} {\bibfnamefont {A.~M.}\ \bibnamefont {Jones}}, \bibinfo
  {author} {\bibfnamefont {R.-L.}\ \bibnamefont {Chu}}, \bibinfo {author}
  {\bibfnamefont {J.}~\bibnamefont {Yan}}, \bibinfo {author} {\bibfnamefont
  {D.~G.}\ \bibnamefont {Mandrus}}, \bibinfo {author} {\bibfnamefont
  {C.}~\bibnamefont {Zhang}}, \bibinfo {author} {\bibfnamefont
  {D.}~\bibnamefont {Cobden}}, \bibinfo {author} {\bibfnamefont
  {W.}~\bibnamefont {Yao}}, \ and\ \bibinfo {author} {\bibfnamefont
  {X.}~\bibnamefont {Xu}},\ }\href {\doibase 10.1038/nphys3201} {\bibfield
  {journal} {\bibinfo  {journal} {Nature Physics}\ }\textbf {\bibinfo {volume}
  {11}},\ \bibinfo {pages} {148} (\bibinfo {year} {2015})}\BibitemShut
  {NoStop}%
\bibitem [{\citenamefont {Wo\ifmmode~\acute{z}\else \'{z}\fi{}niak}\ \emph
  {et~al.}(2020)\citenamefont {Wo\ifmmode~\acute{z}\else \'{z}\fi{}niak},
  \citenamefont {Faria~Junior}, \citenamefont {Seifert}, \citenamefont
  {Chaves},\ and\ \citenamefont {Kunstmann}}]{Wozniak2020}%
  \BibitemOpen
  \bibfield  {author} {\bibinfo {author} {\bibfnamefont {T.}~\bibnamefont
  {Wo\ifmmode~\acute{z}\else \'{z}\fi{}niak}}, \bibinfo {author} {\bibfnamefont
  {P.~E.}\ \bibnamefont {Faria~Junior}}, \bibinfo {author} {\bibfnamefont
  {G.}~\bibnamefont {Seifert}}, \bibinfo {author} {\bibfnamefont
  {A.}~\bibnamefont {Chaves}}, \ and\ \bibinfo {author} {\bibfnamefont
  {J.}~\bibnamefont {Kunstmann}},\ }\href {\doibase
  10.1103/PhysRevB.101.235408} {\bibfield  {journal} {\bibinfo  {journal}
  {Phys. Rev. B}\ }\textbf {\bibinfo {volume} {101}},\ \bibinfo {pages}
  {235408} (\bibinfo {year} {2020})}\BibitemShut {NoStop}%
\bibitem [{\citenamefont {Molas}\ \emph
  {et~al.}(2019{\natexlab{b}})\citenamefont {Molas}, \citenamefont
  {Slobodeniuk}, \citenamefont {Nogajewski}, \citenamefont {Bartos},
  \citenamefont {Bala}, \citenamefont {Babi\ifmmode~\acute{n}\else
  \'{n}\fi{}ski}, \citenamefont {Watanabe}, \citenamefont {Taniguchi},
  \citenamefont {Faugeras},\ and\ \citenamefont
  {Potemski}}]{Molas2019Spectrum}%
  \BibitemOpen
  \bibfield  {author} {\bibinfo {author} {\bibfnamefont {M.~R.}\ \bibnamefont
  {Molas}}, \bibinfo {author} {\bibfnamefont {A.~O.}\ \bibnamefont
  {Slobodeniuk}}, \bibinfo {author} {\bibfnamefont {K.}~\bibnamefont
  {Nogajewski}}, \bibinfo {author} {\bibfnamefont {M.}~\bibnamefont {Bartos}},
  \bibinfo {author} {\bibfnamefont {L.}~\bibnamefont {Bala}}, \bibinfo {author}
  {\bibfnamefont {A.}~\bibnamefont {Babi\ifmmode~\acute{n}\else
  \'{n}\fi{}ski}}, \bibinfo {author} {\bibfnamefont {K.}~\bibnamefont
  {Watanabe}}, \bibinfo {author} {\bibfnamefont {T.}~\bibnamefont {Taniguchi}},
  \bibinfo {author} {\bibfnamefont {C.}~\bibnamefont {Faugeras}}, \ and\
  \bibinfo {author} {\bibfnamefont {M.}~\bibnamefont {Potemski}},\ }\href
  {\doibase 10.1103/PhysRevLett.123.136801} {\bibfield  {journal} {\bibinfo
  {journal} {Phys. Rev. Lett.}\ }\textbf {\bibinfo {volume} {123}},\ \bibinfo
  {pages} {136801} (\bibinfo {year} {2019}{\natexlab{b}})}\BibitemShut
  {NoStop}%
\end{thebibliography}%


\begin{thebibliography}{99}
	
	\bibitem{GomezSI}
	A. Castellanos-Gomez, M. Buscema, R. Molenaar, V. Singh, L. Janssen, H. S. J. van der Zant, and G. A. Steele, 2D Materials {\bf 1}, 011002 (2014).

	\bibitem{VASPSI}	
	G. Kresse and J. Furthmüller, Phys. Rev. B {\bf 54}, 11169 (1996).

	\bibitem{PAWSI}
	G. Kresse and D. Joubert, Phys. Rev. B {\bf 59}, 1758 (1999).

	\bibitem{PBEsolSI}
	J. P. Perdew, A. Ruzsinszky, G. I. Csonka, O. A. Vydrov, G. E. Scuseria, L. A. Constantin, X. Zhou, and K. Burke, Phys. Rev. Lett. {\bf 100}, 136406 (2008).

	\bibitem{SchutteSI}
	W. Schutte, J. De Boer, and F. Jellinek, Journal of Solid State Chemistry {\bf 70}, 207 (1987).

	\bibitem{PhonopySI}
	A. Togo and I. Tanaka, Scr. Mater. {\bf 108}, 1 (2015).

	\bibitem{ParlinskiSI}
	K. Parlinski, Z. Q. Li, and Y. Kawazoe, Phys. Rev. Lett. {\bf 78}, 4063 (1997).

	\bibitem{DFPTSI}
	M. Gajdoš, K. Hummer, G. Kresse, J. Furthmüller, and F. Bechstedt, Phys. Rev. B {\bf 73}, 045112 (2006).

	\bibitem{BarboneSI}
	M. Barbone, A. R. P. Montblanch, D. M. Kara, C. Palacios-Berraquero, A. R. Cadore, D. De Fazio, B. Pingault, E. Mostaani, H. Li, B. Chen, K. Watanabe, T. Taniguchi, S. Tongay, G. Wang, A. C. Ferrari, and M. Atatüre, Nature Communications {\bf 9}, 3721 (2018).

	\bibitem{ChenSI}
	S.-Y. Chen, T. Goldstein, T. Taniguchi, K. Watanabe, and J. Yan, Nature Communications {\bf 9}, 3717 (2018).

	\bibitem{LiSI}
	Z. Li, T. Wang, Z. Lu, C. Jin, Y. Chen, Y. Meng, Z. Lian, T. Taniguchi, K. Watanabe, S. Zhang, D. Smirnov, and S.-F. Shi, Nature Communications {\bf 9}, 3719 (2018).

	\bibitem{KoperskiSI}
	M. Koperski, M. R. Molas, A. Arora, K. Nogajewski, M. Bartos, J. Wyzula, D. Vaclavkova, P. Kossacki, and M. Potemski, 2D Materials {\bf 6}, 015001 (2019).

	\bibitem{RobertSI}
	C. Robert, H. Dery, L. Ren, D. van Tuan, E. Courtade, M. Yang, B. Urbaszek, D. Lagarde, K. Watanabe, T. Taniguchi, T. Amand, and X. Marie, arXiv:2008.07464 [cond-mat.mtrl-sci] (2020).

	\bibitem{ZinkiewiczSI}
	M. Zinkiewicz, A. O. Slobodeniuk, T. Kazimierczuk, P. Kapuściński, K. Oreszczuk, M. Grzeszczyk, M. Bartos, K. Nogajewski, K. Watanabe, T. Taniguchi, C. Faugeras, P. Kossacki, M. Potemski, A. Babiński, and M. R. Molas, Nanoscale {\bf 12}, 18153 (2020).

	\bibitem{StierSI}
	A. V. Stier, K. M. McCreary, B. T. Jonker, J. Kono, and S. A. Crooker, Nature Communications {\bf 7}, 10643 (2016).

\end{thebibliography}

\newpage
\onecolumngrid
\setcounter{figure}{0}
\setcounter{section}{0}
\renewcommand{\thefigure}{S\arabic{figure}}
\renewcommand{\thesection}{S\arabic{section}}

\begin{center}
	{\large{ {\bf Supporting Information:  \\ Excitonic complexes in $n$-doped WS$_2$  monolayer}}}
	\vskip0.5\baselineskip{M. Zinkiewicz,{$^{1}$} T. Wo\'zniak,{$^{2}$} T. Kazimierczuk,{$^{1}$} P. Kapu\'sci\'nski,{$^{3,4}$} K.~Oreszczuk,{$^{1}$} \linebreak[4] M.~Grzeszczyk,{$^{1}$} M. Bartos,{$^{3,5}$} K. Nogajewski,{$^{1}$} K.~Watanabe,{$^{6}$} T. Taniguchi,{$^{7}$} \linebreak[4] C. Faugeras,{$^{3}$} P. Kossacki,{$^{1}$} M. Potemski,{$^{1,3}$} A. Babi\'nski,{$^{1}$} and M. R. Molas{$^{1}$}}		
	\vskip0.5\baselineskip{{\small $^{1}$\textit{Institute of Experimental Physics, Faculty of Physics, University of Warsaw, ul. Pasteura 5, 02-093 Warsaw, Poland} \\$^{2}$\textit{Department of Semiconductor Materials Engineering, Wrocław University of Science and Technology, Wybrzeże Wyspiańskiego 27, 50-370 Wrocław, Poland} \\$^{3}$\textit{Laboratoire National des Champs Magn\'etiques Intenses, CNRS-UGA-UPS-INSA-EMFL, 25, avenue des Martyrs, 38042 Grenoble, France} \\$^{4}$\textit{Department of Experimental Physics, Wrocław University of Science and Technology, ul. Wybrzeże Wyspiańskiego 27, 50-370 Wrocław, Poland} \\$^{5}$\textit{Central European Institute of Technology, Brno University of Technology,  Purky\v{n}ova 656/123, 612 00 Brno, Czech Republic} \\$^{6}$\textit{Research Center for Functional Materials, National Institute for Materials Science, 1-1 Namiki, Tsukuba 305-0044, Japan} \\$^{7}$\textit{International Center for Materials Nanoarchitectonics, National Institute for Materials Science, 1-1 Namiki, Tsukuba 305-0044, Japan}}}
\end{center}

\section{S\lowercase{ample}}
The studied sample is composed of WS$_2$ ML encapsulated in hBN flakes and supported by a bare Si substrate. The structure was obtained by two-stage polydimethylsiloxane (PDMS)-based \cite{GomezSI} mechanical exfoliation of WS$_2$ and hBN bulk crystals. A bottom layer of hBN in the hBN/WS$_2$/hBN heterostructure was created in the course of a non-deterministic exfoliation. The assembly of the hBN/WS$_2$/hBN heterostructure was realized via succesive dry transfers of WS$_2$ ML and capping hBN flake from PDMS stamps onto the bottom hBN layer.

\section{E\lowercase{xperimental setups}}
Low-temperature micro-magneto-PL experiments are performed in the Voigt, Faraday and tilted geometries, $i.e.$ magnetic field oriented parallel, perpendicular, and 45$^\circ$ with respect to ML's plane. Measurements (spatial resolution $\sim$2~$\mu$m) were carried out with the aid of two systems: a split-coil superconducting magnet and a resistive solenoid producing fields up to 10~T and 30~T using a free-beam-optics arrangement and an optical-fiber-based insert, respectively. The sample was placed on top of a $x$-$y$-$z$ piezo-stage kept at $T$=10~K or $T$=4.2~K and was excited using a laser diode with 532~nm or 515~nm wavelength (2.33 eV or 2.41 eV photon energy). The emitted light was dispersed with a 0.5~m long monochromator and detected with a charge coupled device (CCD) camera. For measurements up to 10~T, the circular polarizations of the emissions were analyzed using a set of polarizers and a $\lambda$/4 waveplate placed directly in front of the spectrometer. In tilted-field configurations in high magnetic fields, the combination of a quarter wave plate and a linear polarizer placed in the insert were used to analyse the circular polarization of signals (the measurements were performed with a fixed circular polarization, whereas reversing the direction of magnetic field yields the information corresponding to the other polarization component due to time-reversal symmetry). Note that the excitation power for experiments performed in magnetic fields up to 30~T was adjusted based on the comparison of the measured PL spectrum and the one obtained under excitation of laser with 532~nm.


\section{T\lowercase{heoretical calculations}}
First principles calculations were performed with the use of \textit{Vienna Ab Initio Simulation Package} (VASP) \cite{VASPSI} and the Projector Augmented Wave method \cite{PAWSI}. After testing several exchange-correlation functionals, van der Waals corrections and functionals, the parametrization of Perdew-Burke-Ernzerhof revised for solids (PBEsol) \cite{PBEsolSI} was used for geometry optimization, as yielding the best agreement with experimental lattice constant and layer thickness \cite{SchutteSI}. Atomic positions and lattice constants were optimized with $10^{-5}$ eV/\AA ~and 0.1 kbar precision. Energy cutoff of 400 eV, a $12\times12\times1$ k-mesh and 15 {\AA} of vacuum in the vertical direction of unit cell were chosen after careful convergence tests. Phonon band structure calculations were performed using Phonopy package \cite{PhonopySI}, which implements the finite displacement method to obtain the interatomic force constants \cite{ParlinskiSI}. A $3\times3\times1$ supercell was found sufficient to yield converged phonon energies at $\Gamma$ and K points. The orbital angular momenta of bands were obtained from the wave function derivatives that are calculated within density functional perturbation theory \cite{DFPTSI}. 

\section{E\lowercase{xcitation power evolutions of excitonic emissions}}
\begin{figure}[h]
		\centering
		\includegraphics[width=0.5\linewidth]{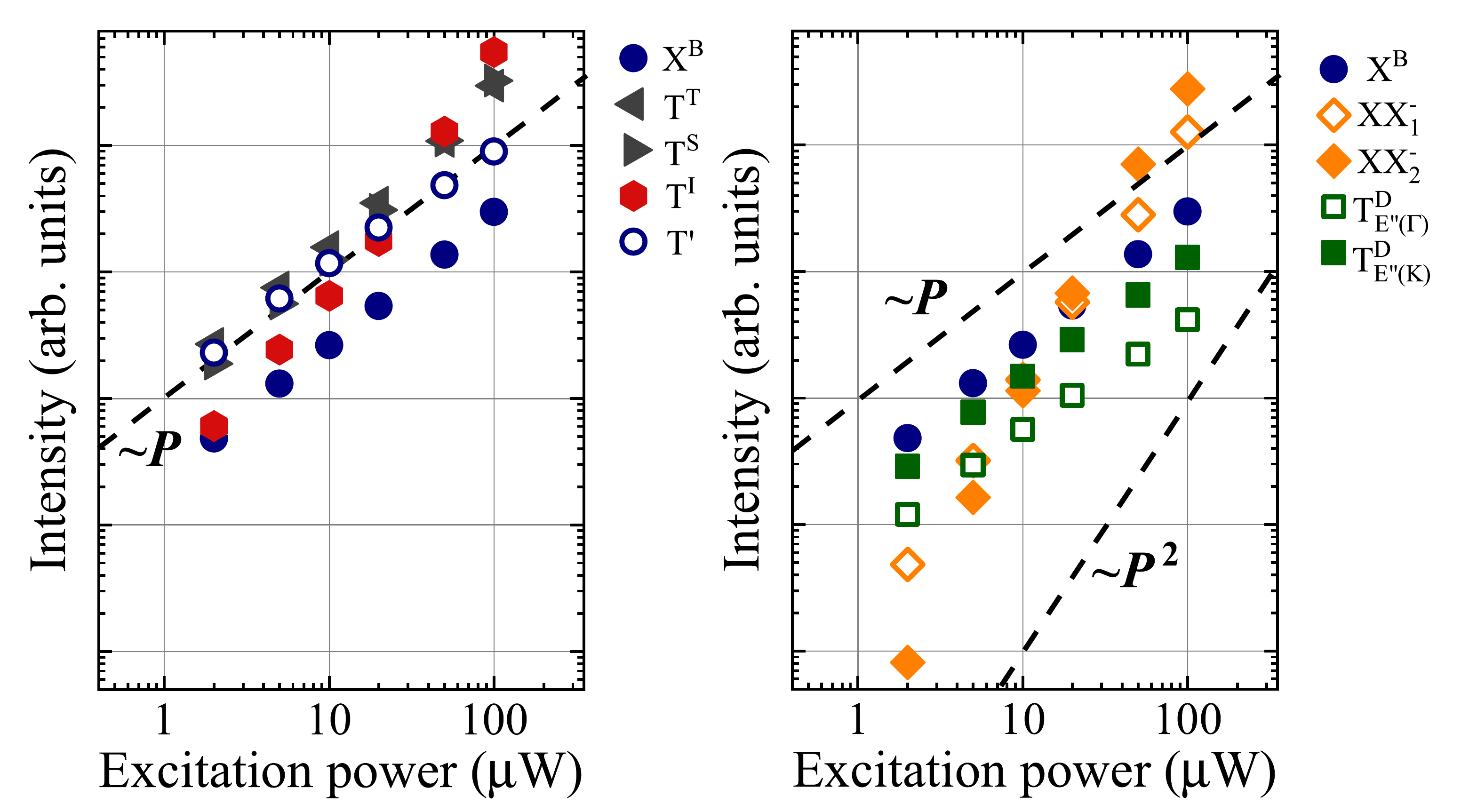}%
		\caption{The intensity evolution of the emission features with excitation power in the log-log plot.  The dashed black line indicates the linear and quadratic behaviours as a guide to the eye.}
		\label{fig:power}
	\end{figure}

In order to verify the assignment of different investigated excitonic complexes, we measured excitation power dependency of the PL spectra. Fig.~\ref{fig:power} demonstrates the integrated intensity of excitons as a function of excitation powers. As can be appreciated, most of the complexes, $i.e.$ X$^\textrm{B}$, T$^\textrm{T}$, T$^\textrm{S}$, T$^\textrm{I}$, T', T$^\textrm{D}_\textrm{E''($\Gamma$)}$, and T$^\textrm{D}_\textrm{E''(K)}$, are characterized by nearly linear dependence, while the intensity growth of both the XX$^-_1$ and XX$^-_2$ lines are described by superlinear evolution. These types of power dependence are typical for excitonic complexes composed of a single electron-hole ($e$-$h$) pair or by two $e$-$h$ pairs~\cite{BarboneSI,ChenSI,LiSI}.

\section{$g$\lowercase{-factors of excitonic complexes}}
\begin{figure}[h]
		\centering
		\includegraphics[width=0.6\linewidth]{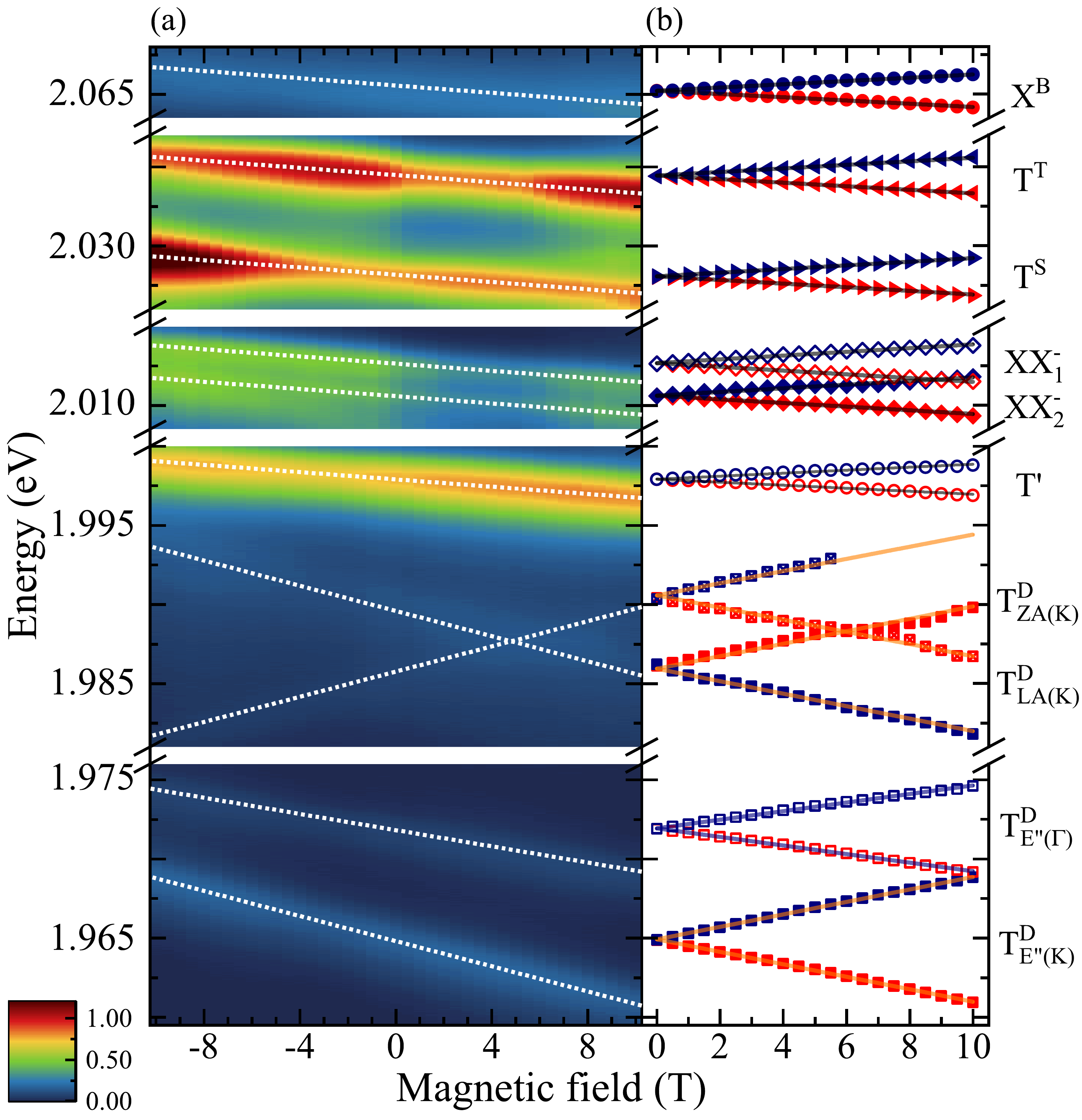}%
		\caption{(a) False-color map of the PL response as a function of $B_\perp$. Note that the positive and negative values of magnetic fields correspond to $\sigma^\pm$ polarizations of detection. The intensity scale is logarithmic. White dashed lines superimposed on the investigated transitions are guides to the eyes. (b) Transition energies of the $\sigma^{+/-}$ (red/blue points) components of different excitonic complexes lines as a function of the~out-of-plane magnetic field. The solid lines represent fits according to Eq.~\ref{eq:zeeman}. Three groups of $g$-factors with absolute values close to 4, 9 and 13 are indicated with different colours black, navy and orange respectively. }
		\label{fig:mapa}
	\end{figure}

Fig.~\ref{fig:mapa}(a) illustrates the false-color map of the magneto-photoluminescence spectra measured in magnetic fields up to 10~T oriented perpendicular to ML's plane under power excitation of 20 $\mu$W, which allows us to investigate the g-factors of negatively charged biexcitons (XX$^-_1$ and XX$^-_2$). Upon application of an out-of-plane magnetic field, the excitonic emissions split into two circularly polarized components due to the excitonic Zeeman effect~\cite{KoperskiSI}. Their energies evolutions ($E(B)$) in external out-of-plane magnetic fields ($B_\perp$) can be described as:
\begin{equation}
E(B)=E_0\pm\frac{1}{2} g \mu_\mathrm{B} B_\perp,
\label{eq:zeeman}
\end{equation}
where $E_0$ is the energy of the transition at zero field, $g$ denotes the $g$-factor of the considered excitonic complex and $\mu_{B}$ is the Bohr magneton. The fitting results of Eq.~\ref{eq:zeeman} to the experimental results are presented in Fig.~\ref{fig:mapa}(b). As can be seen, for all the excitonic complexes, the observed evolutions can be described by the aforementioned formula. We found that the obtained values of $g$-factors can be organized in three groups: $\sim$4 (black fitted lines), $\sim$9 (navy fitted lines), and $\sim$13 (orange fitted lines). The $g$-factors value for each excitonic complex as well as the origin of thee groups are presented in the main article.

\newpage

\section{C\lowercase{onduction and valence bands $g$-factors}}

\begin{figure}[t]
		\centering
		\includegraphics[width=1\linewidth]{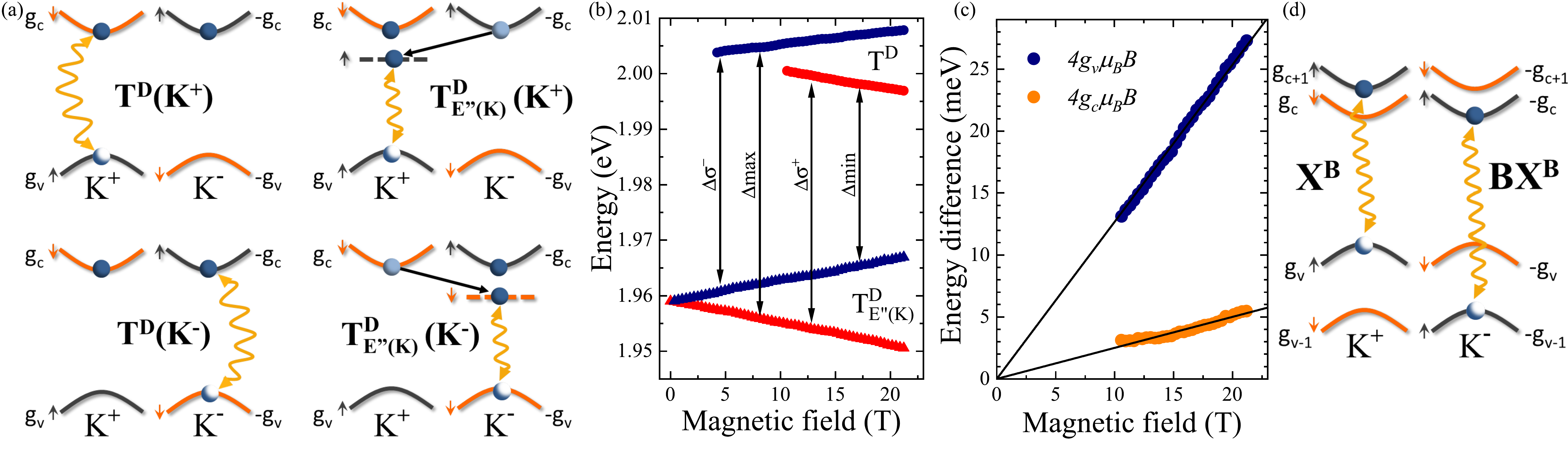}%
		\caption{(a) Schematic illustration of possible configurations for spin-forbidden negative dark trion (T$^\textrm{D}$) negatively charged dark exciton and its phonons replica (T$^\textrm{D}_{\textrm{E"(K)}}$) for both valleys denoted as K$^+$ and K$^-$ in the bracket. (b) Transition energies of the $\sigma^{+/-}$ (red/blue points) components of the T$^\textrm{D}$ and T$^\textrm{D}_{\textrm{E"(K)}}$ transitions as a function of the~out-of-plane magnetic field. Black arrows show energy differences between optical transition marked as $\Delta min$, $\Delta max$, $\Delta \sigma^+$ and $\Delta \sigma^-$. (c) Magnetic field evolution of energy differences described by Eq.~\ref{eq:gv}. Black solid lines represent linear fits to the designated points. (d) Schematic illustration of possible configurations for the bright A (X$^\textrm{B}$) and B (BX$^\textrm{B}$) excitons formed in the K$^+$ and K$^-$ points, respectively.}
		\label{fig:factors}
	\end{figure}

Recently, a method for determination of single subbands $g$-factors in the WSe$_2$ monolayers was proposed in Ref.~\citenum{RobertSI}. As low temperature PL spectra of both the WSe$_2$ and WS$_2$ monolayers comprise of several emission lines including so-called phonon replicas, we have adapted this method to obtain $g$-factors of single conduction and valence subbands in WS$_2$ ML. Within this approach, we compare energy evolution of spin-forbidden dark trion (T$^\textrm{D}$) and its phonon replica (T$^\textrm{D}_{\textrm{E"(K)}}$) in out-of-plane magnetic fields ($B_\perp$). Particularly, the energy difference between their circularly-polarized components ($\sigma^\pm$) corresponding to emissions from different valleys (K$^+$ or K$^-$) is analysed, see Fig.~\ref{fig:factors}(a). Fig.~\ref{fig:factors}(b) shows the magnetic field evolution of the $\sigma^\pm$ components of the T$^\textrm{D}$ and T$^\textrm{D}_{\textrm{E"(K)}}$ lines. Note that the field dependence of T$^\textrm{D}$ emission was measured in the tilted configuration of the sample by 45$^o$ with respect to the applied magnetic field direction, which gives both in-plane and out-of-plane components of magnetic fields, see Ref.~\citenum{ZinkiewiczSI} for details. There are four energy distances marked in Fig.~\ref{fig:factors}(b) by black vertical lines: $\Delta\sigma^+$, $\Delta\sigma^-$, $\Delta min$ and $\Delta max$. With help of schematic illustration shown in Fig.~\ref{fig:factors}(a), it follows that they depend only on single subband $g$-factors and phonon energy and are equal to:

\begin{align}
\Delta \sigma^+=\textrm{T}^\textrm{D}(\textrm{K}^+)-\textrm{T}^\textrm{D}_{\textrm{E}"(\textrm{K})}(\textrm{K}^+)=\textrm{E}"(\textrm{K})+2g_{c}\mu_BB_\perp,\nonumber\\ 
\Delta \sigma^-=\textrm{T}^\textrm{D}(\textrm{K}^-)-\textrm{T}^\textrm{D}_{\textrm{E}"(\textrm{K})}(\textrm{K}^-)=\textrm{E}"(\textrm{K})-2g_{c}\mu_BB_\perp,\nonumber \\    
\Delta min=\textrm{T}^\textrm{D}(\textrm{K}^+)-\textrm{T}^\textrm{D}_{\textrm{E}"(\textrm{K})}(\textrm{K}^-)=\textrm{E}"(\textrm{K})-2g_{v}\mu_BB_\perp, \nonumber    \\ 
\Delta max=\textrm{T}^\textrm{D}(\textrm{K}^-)-\textrm{T}^\textrm{\textrm{D}}_{\textrm{E}"(\textrm{K})}(\textrm{K}^+)=\textrm{E}"(\textrm{K})+2g_{v}\mu_BB_\perp,
\label{eq:delty}
\end{align}
where $\textrm{T}^\textrm{D}(\textrm{K}^\pm)$ and $\textrm{T}^\textrm{D}_{\textrm{E}"(\textrm{K})}(\textrm{K}^\pm)$ represent the magnetic field evolutions of $\sigma^\pm$ components of spin forbidden dark trion ($\textrm{T}^\textrm{D}$) and its E" phonon replica ($\textrm{T}^\textrm{D}_{\textrm{E}"(\textrm{K})}$). E"(K) is energy of E" phonon at K point. $g_c$ and $g_v$ are the $g$-factors of the lowest conduction band $g_{c}$ and the highest valence band $g_v$, while $\mu_B$ and $B_\perp$ are related to the Bohr's magneton and the applied out-of-plane magnetic field.

Further transformations of the aforementioned equations lead to establishing equations which allow determination of $g_{c}$ and $g_v$ values:

\begin{align}
4g_{c}\mu_BB=\Delta \sigma^-- \Delta \sigma^+,\nonumber\\ 
4g_{v}\mu_BB=\Delta max-\Delta min,
\label{eq:gv}
\end{align}

Fig.~\ref{fig:factors}(c) demonstrates experimentally obtained evolutions with linear fits marked with solid black lines. Obtained values are equal to 1.1 and 5.5 for $g_{c}$ and $g_v$, respectively. That is in a good agreement with theoretical calculations, which are discussed widely in the main article.

Knowing values of $g_{c}$ and $g_v$, we are able to determine the $g$-factor of top conduction band subband ($g_{c+1}$) from previously obtained $g$-factor of the bright A exciton, $i.e.$ X$^\textrm{B}$ line, with the aid of formula: $g_{\textrm{X}^\textrm{B}}=2(g_{c+1}-g_v)$. We extracted that $g_{c+1}$ value is equal to 3.7. 

Finally, taking $g$-factor of the bright B exciton  ($g_{\textrm{BX}^\textrm{B}})$ from Ref.~\citenum{StierSI}, it is possible to determine the $g$-factor of the bottom valence subband ($g_{v-1}$) using formula: $g_{\textrm{BX}^\textrm{B}}=2(g_{c}-g_{v-1})$. We found that $g_{v-1}$ value is equal to 2.9. 

\end{document}